 \pdfoutput=1
\documentclass[english]{article}
\usepackage[LGR,T1]{fontenc}
\usepackage[latin9]{inputenc}
\usepackage[a4paper]{geometry}
\usepackage{float}
\usepackage{amsmath}
\usepackage{graphicx}
\usepackage{setspace}
\usepackage[authoryear]{natbib}
\makeatletter

\DeclareRobustCommand{\greektext}{%
  \fontencoding{LGR}\selectfont\def\encodingdefault{LGR}}
\DeclareRobustCommand{\textgreek}[1]{\leavevmode{\greektext #1}}
\ProvideTextCommand{\~}{LGR}[1]{\char126#1}

\newcommand{\lyxmathsym}[1]{\ifmmode\begingroup\def\b@ld{bold}
  \text{\ifx\math@version\b@ld\bfseries\fi#1}\endgroup\else#1\fi}

\providecommand{\tabularnewline}{\\}
\floatstyle{ruled}
\newfloat{algorithm}{tbp}{loa}
\providecommand{\algorithmname}{Algorithm}
\floatname{algorithm}{\protect\algorithmname}

\newcommand{\lyxaddress}[1]{
\par {\raggedright #1
\vspace{1.4em}
\noindent\par}
}

\usepackage{footmisc}

\makeatother

\usepackage{babel}
\begin{document}

\title{\textbf{Angle-Based Models for Ranking Data} }

\author{Hang Xu$^{1}$, Mayer Alvo$^{2}$ and Philip L.H. Yu$^{1}$}
\maketitle

\lyxaddress{$^{1}$Department of Statistics and Actuarial Science, The University
of Hong Kong, Hong Kong }
\begin{singlespace}

\lyxaddress{$^{2}$Department of Mathematics and Statistics, University of Ottawa,
Canada }
\end{singlespace}
\begin{abstract}
A new class of general exponential ranking models is introduced which
we label angle-based models for ranking data. A consensus score vector
is assumed, which assigns scores to a set of items, where the scores
reflect a consensus view of the relative preference of the items.
The probability of observing a ranking is modeled to be proportional
to its cosine of the angle from the consensus vector. Bayesian variational
inference is employed to determine the corresponding predictive density.
It can be seen from simulation experiments that the Bayesian variational
inference approach not only has great computational advantage compared
to the traditional MCMC, but also avoids the problem of overfitting
inherent when using maximum likelihood methods. The model also works
when a large number of items are ranked which is usually an NP-hard
problem to find the estimate of parameters for other classes of ranking
models. Model extensions to incomplete rankings and mixture models
are also developed. Real data applications demonstrate that the model
and extensions can handle different tasks for the analysis of ranking
data.

\end{abstract}

\section{Introduction}

\noindent Ranking data are often encountered in practice when judges
(or individuals) are asked to rank a set of $t$ items, which may
be political goals, candidates in an election, types of food, etc..
We see examples in voting and elections, market research and food
preference just to name a few. 

\citet{alvo1991balanced} considered tests of hypotheses related to
problems of trend and independence using only the ranks of the data.
In another direction, the interest may be in modeling the ranking
data. Some of these models are: (i) order-statistics models \citep{Thurstone1927a,yu2000bayesian},
(ii) distance-based models \citep{critchlow1991probability,lee2012mixtures},
(iii) paired-comparison models \citep{Mallows1957}, and (iv) multistage
models \citep{Fligner1988}. A more comprehensive discussion on these
probability ranking models can be found in the book by \citet{alvo2014statistical}.
However, some of these models cannot handle the situation in which
the number of items being ranked is large, nor when incomplete rankings
exist in the data. For distance-based models: (i) there is no closed-form
for the normalizing constants for Spearman distances and (ii) the
modal ranking is discrete over a finite space of $t!$ dimensions
and searching for it will be time consuming when the number of items,
$t$, becomes large. 

In this article, we first propose a new class of general exponential
ranking models called angle-based models for the distribution of rankings.
We assume a consensus score vector $\boldsymbol{\theta}$ which assigns
scores to the items, where the scores reflect a consensus view of
the relative preference of the items. The probability of observing
a ranking is proportional to the cosine of the angle from the consensus
score vector. The distance-based model with Spearman distance can
be seen as a special case of our model. Unlike the Spearman distance-based
model, we obtain a very good approximation of the normalizing constant
of our angle-based model. Note that this approximation allows us to
have the explicit form in the first or second derivative of normalizing
constant which can facilitate the computation of the ranking probabilities
under the model. 

For the parameter estimation of the model, we first place a joint
Gamma-von Mises-Fisher prior distribution on the parameter. We describe
several mathematical difficulties incurred in determining the resulting
posterior distribution and propose to make use of the variational
inference method. From the simulation experiments, it can be seen
that the Bayesian variational inference approach not only has great
computational advantage compared to the traditional Markov Chain Monte
Carlo (MCMC), but also avoids the over-fitting problem in maximum
likelihood estimation (MLE). Our model also works when the number
of items being ranked is large, while it is usually an NP-hard problem
to obtain the parameter estimates for other classes of ranking models.
Model extensions to the incomplete rankings and mixture model are
also discussed. From the simulations and applications, it can be seen
that our extensions can handle well incomplete rankings as well as
the clustering and classification tasks for ranking data.

The article is organized as follows. Section 2 introduces the angle-based
model as well as the Bayesian MCMC approach. In Section 3, we describe
the method of variational inference for our model and derive the predictive
density of a new ranking. In Section 4, we consider model extensions
to incomplete rankings and mixture models for clustering and classification.
In Section 5, we describe several simulation experiments whereas in
Section 6, the methodology is then applied to real data sets including
a sushi data set, ranking data from the American Psychological Association
(APA) presidential election of 1980 and a breast cancer gene expressions
dataset. We conclude with a discussion in Section 7.

\section{Angle-Based Models}

\subsection{Model setup}

\noindent A ranking $\boldsymbol{R}$ represents the order of preference
with respect to a set of items. In ranking $t$ items, labeled $1,\ldots,t$,
a ranking $\boldsymbol{R}=(R(1),\ldots,R(t))^{T}$ is a mapping function
from $1,...,t$ to ranks $1,...,t$, where $R(2)=3$ means that item
2 is ranked third and $R^{-1}(3)=2$ means that the item ranked third
is item 2. It will be more convenient to standardize the rankings
as:
\[
\boldsymbol{y}=\frac{\boldsymbol{R}-\frac{t+1}{2}}{\sqrt{\frac{t(t^{2}-1)}{12}}},
\]
where $\boldsymbol{y}$ is the $t\times1$ vector with $\left\Vert \boldsymbol{y}\right\Vert =1$.

\begin{figure}[h]
\begin{centering}
\includegraphics[scale=0.8]{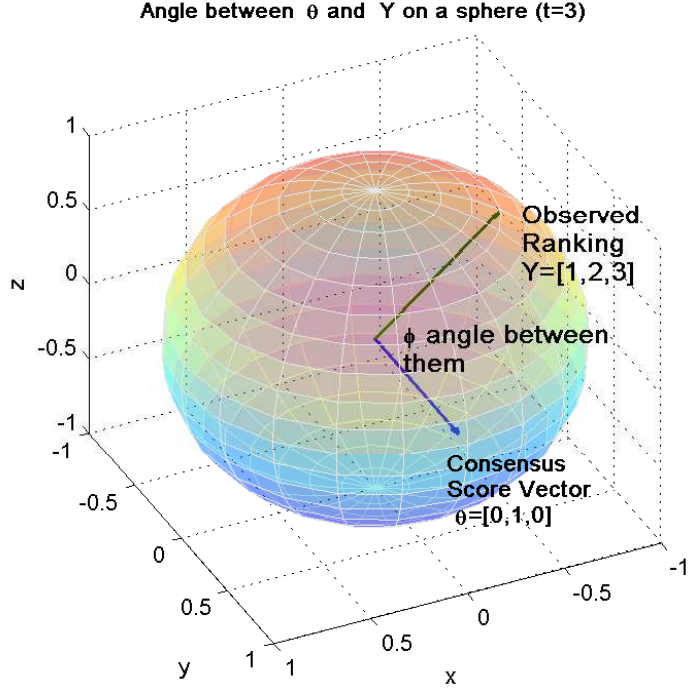}
\par\end{centering}
\caption{\label{fig:Illustration-for-the-angle}Illustration for the angle
between the consensus score vector $\boldsymbol{\theta}=(0,1,0)^{T}$
and the standardized observation of $\left(1,2,3\right)^{T}$ on the
sphere when $t=3.$}
\end{figure}

We consider the following ranking model:
\[
p(\boldsymbol{y}|\kappa,\boldsymbol{\theta})=C(\kappa,\boldsymbol{\theta})\exp\left\{ \kappa\boldsymbol{\theta}^{T}\boldsymbol{y}\right\} ,
\]
where the parameter $\boldsymbol{\theta}$ is a $t\times1$ vector
with $\left\Vert \boldsymbol{\theta}\right\Vert =1$, parameter $\kappa\geq0$,
and $C(\kappa,\boldsymbol{\theta})$ is the normalizing constant.
In the case of the distance-based models \citep{alvo2014statistical},
the parameter $\boldsymbol{\theta}$ can be viewed as if a modal ranking
vector. In fact, if $\boldsymbol{R}$ and $\boldsymbol{\pi}_{0}$
represent an observed ranking and the modal ranking of $t$ items
respectively, then the probability of observing $\boldsymbol{R}$
under the Spearman distance-based model is proportional to
\begin{eqnarray*}
\exp\left\{ -\lambda\left(\frac{1}{2}\sum_{i=1}^{t}\left(R\left(i\right)-\boldsymbol{\pi}_{0}\left(i\right)\right)^{2}\right)\right\}  & = & \exp\left\{ -\lambda\left(\frac{t\left(t+1\right)\left(2t+1\right)}{12}-\boldsymbol{\pi}_{0}^{T}\boldsymbol{R}\right)\right\} \\
 & \propto & \exp\left\{ \kappa\boldsymbol{\theta}^{T}\boldsymbol{y}\right\} ,
\end{eqnarray*}
where $\kappa=\lambda\frac{t(t^{2}-1)}{12}$, and $\boldsymbol{y}$
and $\boldsymbol{\theta}$ are the standardized rankings of $\boldsymbol{R}$
and $\boldsymbol{\pi}_{0}$ respectively. However, the $\boldsymbol{\pi}_{0}$
in the distance-based model is a discrete permutation vector of integers
$\{1,2,\ldots,t\}$ but the $\boldsymbol{\theta}$ in our model is
a real-valued vector, representing a consensus view of the relative
preference of the items from the individuals. Since both $\left\Vert \boldsymbol{\theta}\right\Vert =1$
and $\left\Vert \boldsymbol{y}\right\Vert =1$, the term $\boldsymbol{\theta}^{T}\boldsymbol{y}$
can be seen as $\cos\phi$ where $\phi$ is the angle between the
consensus score vector $\boldsymbol{\theta}$ and the observation
$\boldsymbol{y}$. Figure \ref{fig:Illustration-for-the-angle} illustrates
an example of the angle between the consensus score vector $\boldsymbol{\theta}=(0,1,0)^{T}$
and the standardized observation of $\boldsymbol{R}=\left(1,2,3\right)^{T}$
on the sphere for $t=3$. The probability of observing a ranking is
proportional to the cosine of the angle from the consensus score vector.
The parameter $\kappa$ can be viewed as a concentration parameter.
For small $\kappa$, the distribution of rankings will appear close
to a uniform whereas for larger values of $\kappa$, the distribution
of rankings will be more concentrated around the consensus score vector. 

To compute the normalizing constant $C(\kappa,\boldsymbol{\theta})$,
let $\lyxmathsym{\textgreek{R}}_{t}$ be the set of all possible permutations
of the integers $1,...,t$. Then 
\begin{equation}
\left(C(\kappa,\boldsymbol{\theta})\right)^{-1}=\sum_{\boldsymbol{y}\in\lyxmathsym{\textgreek{R}}_{t}}\exp\left\{ \kappa\boldsymbol{\theta}^{T}\boldsymbol{y}\right\} .\label{eq:normalizing constant}
\end{equation}
Notice that the summation is over $t!$ elements in $\lyxmathsym{\textgreek{R}}_{t}$.
When $t$ is large, say greater than 15, the exact calculation of
the normalizing constant is prohibitive. Using the fact that the set
of $t!$ permutations lie on a sphere in $(t-1)$-space, our model
resembles the continuous von Mises-Fisher distribution, abbreviated
as $vMF(\boldsymbol{x}|\boldsymbol{m},\kappa)$, which is defined
on a $\left(p-1\right)$ unit sphere with mean direction $\boldsymbol{m}$
and concentration parameter $\kappa$:
\[
p(\boldsymbol{x}|\kappa,\boldsymbol{m})=V_{p}(\kappa)\exp(\kappa\boldsymbol{m}^{T}\boldsymbol{x}),
\]
where 
\[
V_{p}(\kappa)=\frac{\kappa^{\frac{p}{2}-1}}{\left(2\pi\right)^{\frac{p}{2}}I_{\frac{p}{2}-1}(\kappa)},
\]
and $I_{\frac{p}{2}-1}(\kappa)$ is the modified Bessel function of
the first kind with order $\frac{p}{2}-1.$ Consequently, we may approximate
the sum in (\ref{eq:normalizing constant}) by an integral over the
sphere. It is shown in Appendix A that 
\[
C(\kappa,\boldsymbol{\theta})\simeq C_{t}(\kappa)=\frac{\kappa^{\frac{t-3}{2}}}{2^{\frac{t-3}{2}}t!I_{\frac{t-3}{2}}(\kappa)\Gamma(\frac{t-1}{2})},
\]
 where $\Gamma(\cdot)$ is the gamma function. Table \ref{tab:The-error-rate-of-NC}
shows the error rate of the approximate log-normalizing constant as
compared to the exact one computed by direct summation. Here, $\kappa$
is chosen to be 0.01 to 2 and $t$ ranges from 3 to 11. Note that
the exact calculation of the normalizing constant for $t=11$ requires
the summation of $11!\approx3.9\times10^{7}$ permutations. The computer
ran out of memory (16GB) beyond $t=11$. This approximation seems
to be very accurate even when $t=3$. The error drops rapidly as $t$
increases. Note that this approximation allows us to approximate the
first and second derivatives of $\log C$ which can facilitate our
computation in what follows.

\begin{figure}[h]
\begin{centering}
\includegraphics[scale=0.6]{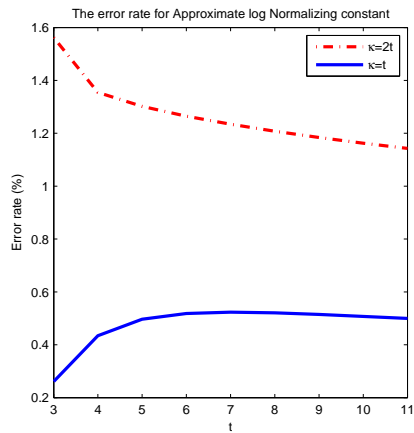}
\par\end{centering}
\caption{\label{fig:The-error-rate-of-NC-1}The error rate of the approximate
log-normalizing constant as compared to the exact one computed by
direct summation for $\kappa=t$ and $\kappa=2t$.}
\end{figure}

Notice that $\kappa$ may grow with $t$ as $\boldsymbol{\theta}^{T}\boldsymbol{y}$
is a sum of $t$ terms. It can be seen from the applications in Section
6 that in one of the clusters for the APA data ($t=5$), $\kappa$
is $7.44$($\approx1.5t$) (see Table 4) while in the gene data ($t=96$),
$\kappa$ is $194.34(\approx2.0t)$ (see Table 5). We thus compute
the error rate for $\kappa=t$ and $\kappa=2t$ as shown in Figure
\ref{fig:The-error-rate-of-NC-1}. It is found that the approximation
is still accurate with error rate of less than 0.5\% for $\kappa=t$
and is acceptable for large $t$ when $\kappa=2t$ as the error rate
decreases in $t$. The von Mises-Fisher distribution was used to model
compositional data by \citet{hornik2014movmf} who also provide different
approaches for estimating $\kappa$ efficiently. 

\begin{table}[h]
\begin{centering}
{\scriptsize{}\hspace*{-1cm}}%
\begin{tabular}{cccccccccc}
\hline 
 & \multicolumn{9}{c}{{\footnotesize{}$t$}}\tabularnewline
\cline{2-10} 
{\footnotesize{}$\kappa$} & {\footnotesize{}3} & {\footnotesize{}4} & {\footnotesize{}5} & {\footnotesize{}6} & {\footnotesize{}7} & {\footnotesize{}8} & {\footnotesize{}9} & {\footnotesize{}10} & {\footnotesize{}11}\tabularnewline
\hline 
{\footnotesize{}$0.01$} & {\footnotesize{}<0.00001\%} & {\footnotesize{}<0.00001\%} & {\footnotesize{}<0.00001\%} & {\footnotesize{}<0.00001\%} & {\footnotesize{}<0.00001\%} & {\footnotesize{}<0.00001\%} & {\footnotesize{}<0.00001\%} & {\footnotesize{}<0.00001\%} & {\footnotesize{}<0.00001\%}\tabularnewline
{\footnotesize{}$0.1$} & {\footnotesize{}<0.00001\%} & {\footnotesize{}<0.00001\%} & {\footnotesize{}<0.00001\%} & {\footnotesize{}<0.00001\%} & {\footnotesize{}<0.00001\%} & {\footnotesize{}<0.00001\%} & {\footnotesize{}<0.00001\%} & {\footnotesize{}<0.00001\%} & {\footnotesize{}<0.00001\%}\tabularnewline
{\footnotesize{}$0.5$} & {\footnotesize{}0.00003\%} & {\footnotesize{}0.00042\%} & {\footnotesize{}0.00024\%} & {\footnotesize{}0.00013\%} & {\footnotesize{}0.00007\%} & {\footnotesize{}0.00004\%} & {\footnotesize{}0.00003\%} & {\footnotesize{}0.00002\%} & {\footnotesize{}0.00001\%}\tabularnewline
{\footnotesize{}$0.8$} & {\footnotesize{}0.00051\%} & {\footnotesize{}0.00261\%} & {\footnotesize{}0.00150\%} & {\footnotesize{}0.00081\%} & {\footnotesize{}0.00046\%} & {\footnotesize{}0.00027\%} & {\footnotesize{}0.00017\%} & {\footnotesize{}0.00011\%} & {\footnotesize{}0.00008\%}\tabularnewline
{\footnotesize{}$1$} & {\footnotesize{}0.00175\%} & {\footnotesize{}0.00607\%} & {\footnotesize{}0.00354\%} & {\footnotesize{}0.00194\%} & {\footnotesize{}0.00110\%} & {\footnotesize{}0.00066\%} & {\footnotesize{}0.00041\%} & {\footnotesize{}0.00027\%} & {\footnotesize{}0.00018\%}\tabularnewline
{\footnotesize{}$2$} & {\footnotesize{}0.05361\%} & {\footnotesize{}0.06803\%} & {\footnotesize{}0.04307\%} & {\footnotesize{}0.02528\%} & {\footnotesize{}0.01508\%} & {\footnotesize{}0.00932\%} & {\footnotesize{}0.00598\%} & {\footnotesize{}0.00398\%} & {\footnotesize{}0.00273\%}\tabularnewline
\hline 
\end{tabular}
\par\end{centering}{\scriptsize \par}
\caption{\label{tab:The-error-rate-of-NC}The error rate of the approximate
log-normalizing constant as compared to the exact one computed by
direct summation. }
\end{table}

\subsection{Maximum likelihood estimation (MLE) of our model\label{subsec:MLE-of-osimple-model}}

\noindent Let $\boldsymbol{Y}=\left\{ \boldsymbol{y}_{1},\ldots,\boldsymbol{y}_{N}\right\} $
be a random sample of $N$ standardized rankings drawn from $p(\boldsymbol{y}|\kappa,\boldsymbol{\theta})$.
The log-likelihood of $\left(\kappa,\boldsymbol{\theta}\right)$ is
then given by
\begin{equation}
L(\boldsymbol{Y}|\kappa,\boldsymbol{\theta})=N\ln C_{t}(\kappa)+\sum_{i=1}^{N}\kappa\boldsymbol{\theta}^{T}\boldsymbol{y}_{i}.\label{eq:log-likelihood-data}
\end{equation}
Maximizing (\ref{eq:log-likelihood-data}) subject to $\left\Vert \boldsymbol{\theta}\right\Vert =1$
and $\kappa\geq0$, we find that the maximum likelihood estimator
of $\boldsymbol{\theta}$ is given by $\hat{\boldsymbol{\theta}}_{MLE}=\frac{\sum_{i=1}^{N}\boldsymbol{y}_{i}}{\left\Vert \sum_{i=1}^{N}\boldsymbol{y}_{i}\right\Vert },$
and $\hat{\kappa}$ is the solution of 
\begin{equation}
A_{t}(\kappa)\equiv\frac{-C_{t}^{'}(\kappa)}{C_{t}(\kappa)}=\frac{I_{\frac{t-1}{2}}\left(\kappa\right)}{I_{\frac{t-3}{2}}\left(\kappa\right)}=\frac{\left\Vert \sum_{i=1}^{N}\boldsymbol{y}_{i}\right\Vert }{N}\equiv r.\label{eq:MLE_alpha}
\end{equation}

A simple approximation to the solution of (\ref{eq:MLE_alpha}) following
\citet{banerjee2005clustering} is given by
\[
\hat{\kappa}_{MLE}=\frac{r(t-1-r^{2})}{1-r^{2}}.
\]
A more precise approximation can be obtained from a few iterations
of Newton's method. Using the method suggested by \citet{sra2012short},
starting from an initial value $\kappa_{0}$, we can recursively update
$\kappa$ by iteration:
\[
\kappa_{i+1}=\kappa_{i}-\frac{A_{t}(\kappa_{i})-r}{1-A_{t}(\kappa_{i})^{2}-\frac{t-2}{\kappa_{i}}A_{t}(\kappa_{i})},\;i=0,1,2,\ldots.
\]

\subsection{Bayesian method with conjugate prior and posterior\label{subsec:Bayesian-method-SIR}}

\noindent Taking a Bayesian approach, we consider the following conjugate
prior for $(\kappa,\boldsymbol{\theta})$ as
\begin{equation}
p(\kappa,\boldsymbol{\theta})\propto\left[C_{t}(\kappa)\right]^{\nu_{0}}\exp\left\{ \beta_{0}\kappa\boldsymbol{m}_{0}^{T}\boldsymbol{\theta}\right\} ,\label{eq:conjugate_prior}
\end{equation}
where $\left\Vert \boldsymbol{m}_{0}\right\Vert =1$, $\nu_{0},\beta_{0}\geq0$.
Given $\boldsymbol{Y}$, the posterior density of $(\kappa,\boldsymbol{\theta})$
can be expressed by
\[
p(\kappa,\boldsymbol{\theta}|\boldsymbol{Y})\propto\exp\left\{ \beta\kappa\boldsymbol{m}^{T}\boldsymbol{\theta}\right\} V_{t}(\beta\kappa)\frac{\left[C_{t}(\kappa)\right]^{N+\nu_{0}}}{V_{t}(\beta\kappa)},
\]
where $\boldsymbol{m}=\left(\beta_{0}\boldsymbol{m}_{\boldsymbol{0}}+\sum_{i=1}^{N}\boldsymbol{y}_{i}\right)\beta^{-1},$
$\beta=\left\Vert \beta_{0}\boldsymbol{m}_{0}+\sum_{i=1}^{N}\boldsymbol{y}_{i}\right\Vert $.
The posterior density can be factorized as 
\begin{equation}
p(\kappa,\boldsymbol{\theta}|\boldsymbol{Y})=p(\boldsymbol{\theta}|\kappa,\boldsymbol{Y})p(\kappa|\boldsymbol{Y}),\label{eq:conjugate_posterior}
\end{equation}
 where $p(\boldsymbol{\theta}|\kappa,\boldsymbol{Y})\sim vMF(\boldsymbol{\theta}|\boldsymbol{m},\beta\kappa)$
and 
\[
p(\kappa|\boldsymbol{Y})\propto\frac{\left[C_{t}(\kappa)\right]^{N+\nu_{0}}}{V_{t}(\beta\kappa)}=\frac{\kappa^{\frac{t-3}{2}(\upsilon_{0}+N)}I_{\frac{t-2}{2}}(\beta\kappa)}{\left[I_{\frac{t-3}{2}}(\kappa)\right]^{\nu_{0}+N}\left(\beta\kappa\right)^{\frac{t-2}{2}}}.
\]
The normalizing constant for $p(\kappa|\boldsymbol{Y})$ is not available
in closed form. \citet{nunez2005bayesian} suggested using a sampling-importance-resampling
(SIR) procedure with a proposal density chosen to be the gamma density
with mean $\hat{\kappa}_{MLE}$ and variance equal to some pre-specified
number such as 50 or 100. However, in a simulation study, it was found
that the choice of this variance is crucially related to the performance
of SIR. An improper choice of variance may lead to slow or unsuccessful
convergence. Also the MCMC method leads to intensive computational
complexity. Furthermore, when the sample size $N$ is large, $\beta\kappa$
can be very large which complicates the computation of the term $I_{\frac{t-2}{2}}\left(\left(\beta\kappa\right)\right)$
in $V_{t}(\beta\kappa).$ Thus the calculation of the weights in the
SIR method will fail when $N$ is large. We conclude that in view
of the difficulties for directly sampling from $p(\kappa|\boldsymbol{Y})$,
it may be preferable to approximate the posterior distribution with
an alternative method known as variational inference (abbreviated
VI from here on).

\section{Variational Inference}

\noindent VI provides a deterministic approximation to an intractable
posterior density through optimization. It has been used in many applications
and tends to be faster than classical methods, such as Markov Chain
Monte Carlo (MCMC) sampling and is easier to scale to large data.
The basic idea behind VI is to first posit a candidate family of densities
and then to select the member of that family which is closest to the
target posterior density as measured by the Kullback-Leibler divergence.
If $q\left(\boldsymbol{Z}\right)$ represents the candidate family
and $p\left(\boldsymbol{Z}|\boldsymbol{Y}\right)$ represents the
target posterior density, the Kullback-Leibler divergence is given
by
\[
KL\left(q|p\right)=E_{q}\left[\ln\frac{q\left(\boldsymbol{Z}\right)}{p\left(Z|\boldsymbol{Y}\right)}\right].
\]
See \citet{blei2017variational} for a more comprehensive discussion
of VI. We first adopt a joint vMF-Gamma distribution as the prior
for $(\kappa,\boldsymbol{\theta})$: 
\begin{align*}
p(\kappa,\boldsymbol{\theta}) & =p(\boldsymbol{\theta}|\kappa)p(\kappa)\\
 & =vMF(\boldsymbol{\theta}|\boldsymbol{m}_{0},\beta_{0}\kappa)Gamma(\kappa|a_{0},b_{0}),
\end{align*}
where $Gamma(\kappa|a_{0},b_{0})$ is the Gamma density function with
shape parameter $a_{0}$ and rate parameter $b_{0}$ (i.e., mean equal
to $\frac{a_{0}}{b_{0}}$), and $p(\boldsymbol{\theta}|\kappa)=vMF(\boldsymbol{\theta}|\boldsymbol{m}_{0},\beta_{0}\kappa)$.
The choice of $Gamma(\kappa|a_{0},b_{0})$ for $p(\kappa)$ is motivated
by the fact that for large values of $\kappa$, $p(\kappa)$ based
on (\ref{eq:conjugate_posterior}) tends to take the shape of a Gamma
density. In fact, for large values of $\kappa$, $I_{\frac{t-3}{2}}(\kappa)\simeq\frac{e^{\kappa}}{\sqrt{2\pi\kappa}},$
and hence $p(\kappa)$ becomes the Gamma density with shape $(\nu_{0}-1)\frac{t-2}{2}+1$
and rate $\nu_{0}-\beta_{0}$:
\[
p(\kappa)\propto\frac{\left[C_{t}(\kappa)\right]^{\nu_{0}}}{V_{t}(\kappa\beta)}\simeq\kappa^{(\nu_{0}-1)\frac{t-2}{2}}\exp(-(\nu-\beta)\kappa).
\]
In a similar vein, \citet{forbes2015fast} used a similar Gamma-based
approximation to develop an algorithm for sampling from the Bessel
exponential posterior distribution for $\kappa.$

Under the usual variational Bayesian methods, all variables are assumed
to be mutually independent. This is known as the mean-field approximation.
However, inspired by the conjugate posterior distribution (\ref{eq:conjugate_posterior}),
we adopt a structural factorization of the variational posterior as
$q(\boldsymbol{\theta},\kappa)=q(\boldsymbol{\theta}|\kappa)q(\kappa)$
which retains the dependency between $\boldsymbol{\theta}$ and $\kappa$.

\subsection{Optimization of the variational distribution\label{subsec:Optimization-of-the-VI}}

\noindent In the variational inference framework, we aim to determine
$q$ so as to minimize the Kullback-Leibler (KL) divergence between
$p(\boldsymbol{\theta},\kappa|\boldsymbol{Y})$ and $q(\boldsymbol{\theta},\kappa)$.
This can be shown to be equivalent to maximizing the evidence lower
bound (ELBO) \citep{blei2017variational}. So the optimization of
the variational factors $q(\boldsymbol{\theta}|\kappa)$ and $q(\kappa)$
is performed by maximizing the evidence lower bound $\mathcal{L}(q)$
with respect to $q$ on the log-marginal likelihood, which in our
model is given by
\begin{align}
\mathcal{L}(q) & =E_{q(\boldsymbol{\theta},\kappa)}\left[\ln\frac{p(\boldsymbol{y}|\kappa,\boldsymbol{\theta})p(\boldsymbol{\theta}|\kappa)p(\kappa)}{q(\boldsymbol{\theta}|\kappa)q(\kappa)}\right]\label{eq:evidence_lower_bound}\\
 & =E_{q(\boldsymbol{\theta},\kappa)}\left[f(\boldsymbol{\theta},\kappa)\right]-E_{q(\boldsymbol{\theta},\kappa)}\left[\ln q(\boldsymbol{\theta}|\kappa)\right]-E_{q(\kappa)}\left[\ln q(\kappa)\right]+constant,\nonumber 
\end{align}
where all the expectations are taken with respect to $q(\boldsymbol{\theta},\kappa)$
and
\begin{align*}
f(\boldsymbol{\theta},\kappa) & =\sum_{i=1}^{N}\kappa\boldsymbol{\theta}^{T}\boldsymbol{y}_{i}+N\left(\frac{t-3}{2}\right)\ln\kappa-N\ln I_{\frac{t-3}{2}}(\kappa)+\kappa\beta_{0}\boldsymbol{m}_{0}^{T}\boldsymbol{\theta}\\
 & +\left(\frac{t-2}{2}\right)\ln\kappa-\ln I_{\frac{t-2}{2}}(\kappa\beta_{0})+(a_{0}-1)\ln\kappa-b_{0}\kappa.
\end{align*}
For fixed $\kappa$, the optimal posterior distribution $\ln q^{*}(\boldsymbol{\theta}|\kappa)$
is $\ln q^{*}(\boldsymbol{\theta}|\kappa)=\kappa\beta_{0}\boldsymbol{m}_{0}^{T}\boldsymbol{\theta}+\sum_{i=1}^{N}\kappa\boldsymbol{\theta}^{T}\boldsymbol{y}_{i}+constant.$
We recognize $q^{*}(\boldsymbol{\theta}|\kappa)$ as a von Mises-Fisher
distribution $vMF(\boldsymbol{\theta}|\boldsymbol{m},\kappa\beta)$
where
\[
\beta=\left\Vert \beta_{0}\boldsymbol{m}_{0}+\sum_{i=1}^{N}\boldsymbol{y}_{i}\right\Vert \;\text{ and }\;\boldsymbol{m}=\left(\beta_{0}\boldsymbol{m}_{0}+\sum_{i=1}^{N}\boldsymbol{y}_{i}\right)\beta^{-1}.
\]

Let $g(\kappa)$ denote the remaining terms in $f(\boldsymbol{\theta},\kappa)$
which only involve $\kappa$:
\begin{align*}
g(\kappa) & =\left[N\left(\frac{t-3}{2}\right)+a_{0}-1\right]\ln\kappa-b_{0}\kappa-N\ln I_{\frac{t-3}{2}}(\kappa)-\ln I_{\frac{t-2}{2}}(\kappa\beta_{0})+\ln I_{\frac{t-2}{2}}(\kappa\beta).
\end{align*}
It is still difficult to maximize $E_{q(\kappa)}\left[g(\kappa)\right]-E_{q(\kappa)}\left[\ln q(\kappa)\right]$
since it involves the evaluation of the expected modified Bessel function.
Follow the similar idea in \citet{taghia2014bayesian}, we first find
a tight lower bound $\underline{g(\kappa)}$ for $g(\kappa)$ so that
\[
\mathcal{L}(q)\geq\underline{\mathcal{L}(q)}=E_{q(\kappa)}\left[\underline{g(\kappa)}\right]-E_{q(\kappa)}\left[\ln q(\kappa)\right]+constant.
\]
From the properties of the modified Bessel function of the first kind,
it is known that the function $\ln I_{\nu}(x)$ is strictly concave
with respect to $x$ and strictly convex relative to $\ln x$ for
all $\nu>0$. Then, we can have the following two inequalities:
\begin{equation}
\ln I_{\nu}(x)\leq\ln I_{\nu}(\bar{x})+\left(\frac{\partial}{\partial x}\ln I_{\nu}(\bar{x})\right)(x-\bar{x}),\label{eq:Ineq-I}
\end{equation}
\begin{equation}
\ln I_{\nu}(x)\geq\ln I_{\nu}(\bar{x})+\left(\frac{\partial}{\partial x}\ln I_{\nu}(\bar{x})\right)\bar{x}(\ln x-\ln\bar{x}).\label{eq:Ineq-II}
\end{equation}
where $\frac{\partial}{\partial x}\ln I_{\nu}(\bar{x})$ is the first
derivative of $\ln I_{\nu}(x)$ evaluated at $x=\bar{x}$. Applying
inequality (\ref{eq:Ineq-I}) for $\ln I_{\frac{t-3}{2}}(\kappa)$
and inequality (\ref{eq:Ineq-II}) for $\ln I_{\frac{t-2}{2}}(\kappa\beta_{0})$,
we have
\begin{align*}
g(\kappa) & \geq\underline{g(\kappa)}=\left[N\left(\frac{t-3}{2}\right)+a_{0}-1\right]\ln\kappa-b_{0}\kappa+\ln I_{\frac{t-2}{2}}(\beta\bar{\kappa})\\
 & +\frac{\partial}{\partial\beta\kappa}\ln I_{\frac{t-2}{2}}(\beta\bar{\kappa})\beta\bar{\kappa}\left(\ln\beta\kappa-\ln\beta\bar{\kappa}\right)-N\ln I_{\frac{t-3}{2}}(\bar{\kappa})\\
 & -N\frac{\partial}{\partial\kappa}\ln I_{\frac{t-3}{2}}(\bar{\kappa})\left(\kappa-\bar{\kappa}\right)-N\ln I_{\frac{t-2}{2}}(\beta_{0}\bar{\kappa})-N\frac{\partial}{\partial\beta_{0}\kappa}\ln I_{\frac{t-2}{2}}(\beta_{0}\bar{\kappa})\beta_{0}\left(\kappa-\bar{\kappa}\right).
\end{align*}
Since the equality holds when $\kappa=\bar{\kappa}$, we see that
the lower bound of $\mathcal{L}(q)$ is tight. Rearranging the terms,
we have the approximate optimal solution as $\ln q^{*}(\kappa)=(a-1)\ln\kappa-b\kappa+constant,$
where 
\begin{equation}
a=a_{0}+N\left(\frac{t-3}{2}\right)+\beta\bar{\kappa}\left[\frac{\partial}{\partial\beta\kappa}\ln I_{\frac{t-2}{2}}(\beta\bar{\kappa})\right],\label{eq:update_postrerior_a}
\end{equation}
\begin{equation}
b=b_{0}+N\frac{\partial}{\partial\kappa}I_{\frac{t-3}{2}}(\bar{\kappa})+\beta_{0}\left[\frac{\partial}{\partial\beta_{0}\kappa}\ln I_{\frac{t-2}{2}}(\beta_{0}\bar{\kappa})\right].\label{eq:Update_posterior_b}
\end{equation}
 We also recognize $q^{*}(\kappa)$ to be a $Gamma(\kappa|a,b)$ with
shape $a$ and rate $b$. The posterior mode $\bar{\kappa}$ obtained
from the previous iteration as:
\begin{equation}
\bar{\kappa}=\begin{cases}
\frac{a-1}{b} & \mbox{if }a>1,\\
\frac{a}{b} & \mbox{otherwise.}
\end{cases}\label{eq:Update_alpha_ba}
\end{equation}
A summary of the algorithm for our estimation is shown in Algorithm
\ref{alg:Bayesian-Estimation-our_model}.

\begin{algorithm}[h]
\textbf{\small{}Input: }{\small{}Scaled $\boldsymbol{Y}=\left\{ \boldsymbol{y}_{1},...,\boldsymbol{y}_{N}\right\} $}{\small \par}

\textbf{\small{}Step 1: Initialization}{\small \par}
\begin{enumerate}
\item {\small{}Set the prior parameters: $\beta_{0}$, $\boldsymbol{m}_{0}$,
$a_{0}$ and $b_{0}$.}{\small \par}
\item {\small{}Calculate the posterior parameters for $q^{*}(\boldsymbol{\theta}|\kappa)$:
$\boldsymbol{m}$, $\beta$.}{\small \par}
\item {\small{}Calculate the initial value of $\bar{\kappa}=\frac{a_{0}}{b_{0}}$.}{\small \par}
\end{enumerate}
\textbf{\small{}Step 2: Optimization of the posterior distribution}{\small \par}

\textbf{\small{}repeat}{\small \par}
\begin{enumerate}
\item {\small{}Update posterior parameter $a$ and $b$ by (\ref{eq:update_postrerior_a})
and (\ref{eq:Update_posterior_b}) respectively.}{\small \par}
\item {\small{}Update $\bar{\kappa}$ by (\ref{eq:Update_alpha_ba}).}{\small \par}
\end{enumerate}
\textbf{\small{}until convergence}{\small \par}

\caption{\label{alg:Bayesian-Estimation-our_model}Bayesian Estimation using
variational inference of our model}

\end{algorithm}

\subsection{\label{subsec:Predictive-Density-of_our_model}Predictive density
of our model}

\noindent We may derive the predictive density for a new standardized
ranking $\tilde{\boldsymbol{y}}$ given the observed data $\boldsymbol{Y}$.
The exact predictive density is given by
\begin{equation}
p(\tilde{\boldsymbol{y}}|\boldsymbol{Y})=\int\int p(\tilde{\boldsymbol{y}}|\kappa,\boldsymbol{\theta})p(\kappa,\boldsymbol{\theta}|\boldsymbol{Y})\,d\kappa d\boldsymbol{\theta}.\label{eq:exact_predictive_density-1}
\end{equation}
We can approximate this density by first replacing the true posterior
distribution with its variational approximation as:
\begin{align}
p(\tilde{\boldsymbol{y}}|\boldsymbol{Y}) & \approx q(\tilde{\boldsymbol{y}}|\boldsymbol{Y})=\int\int p(\tilde{\boldsymbol{y}}|\kappa,\boldsymbol{\theta})q(\boldsymbol{\theta}|\kappa,\boldsymbol{Y})q(\kappa|\boldsymbol{Y})\,d\kappa d\boldsymbol{\theta}\nonumber \\
 & =\int\int p(\tilde{\boldsymbol{y}}|\kappa,\boldsymbol{\theta})vMF(\boldsymbol{\theta}|\boldsymbol{m},\beta\kappa)d\boldsymbol{\theta}Gamma(\kappa|a,b)\,d\kappa\label{eq:posterior_predictiver_density}
\end{align}
where $\kappa$, $\beta$, $a$ and $b$ are the posterior parameters
calculated from our algorithm.

After using a second-order approximation of the Bessel function, the
approximate predictive density of $\tilde{\boldsymbol{y}}$ can be
obtained by:
\[
q(\tilde{\boldsymbol{y}}|\boldsymbol{Y})\approx h(\tilde{\boldsymbol{y}})l(\bar{\kappa})e^{r(\tilde{\boldsymbol{y}})\bar{\kappa}}\bar{\kappa}^{-s(\tilde{\boldsymbol{y}})}\frac{b^{a+\frac{t-1}{2}-1}\Gamma(a+s(\tilde{\boldsymbol{y}})+\frac{t-1}{2}-1)}{\left(b+r(\tilde{\boldsymbol{y}})\right)^{a+s(\tilde{\boldsymbol{y}})+\frac{t-1}{2}-1}\Gamma(a+\frac{t-1}{2}-1)},
\]
where $\eta(\tilde{\boldsymbol{y}})=\left\Vert \tilde{\boldsymbol{y}}+\beta\boldsymbol{m}\right\Vert $
and
\[
s(\tilde{\boldsymbol{y}})=-\eta^{2}(\tilde{\boldsymbol{y}})\bar{\kappa}^{2}\left(\frac{I'_{\frac{t-2}{2}}(\eta(\tilde{\boldsymbol{y}})\bar{\kappa})}{I_{\frac{t-2}{2}}(\eta(\tilde{\boldsymbol{y}})\bar{\kappa})}\right)'+\beta^{2}\bar{\kappa}^{2}\left(\frac{I'_{\frac{t-2}{2}}(\beta\bar{\kappa})}{I_{\frac{t-2}{2}}(\beta\bar{\kappa})}\right)'+\bar{\kappa}^{2}\left(\frac{I'_{\frac{t-3}{2}}(\bar{\kappa})}{I_{\frac{t-3}{2}}(\bar{\kappa})}\right)',
\]
\[
r(\tilde{\boldsymbol{y}})=\frac{s(\tilde{\boldsymbol{y}})}{\bar{\kappa}}-\eta(\tilde{\boldsymbol{y}})\frac{I'_{\frac{t-2}{2}}(\eta(\tilde{\boldsymbol{y}})\bar{\kappa})}{I_{\frac{t-2}{2}}(\eta(\tilde{\boldsymbol{y}})\bar{\kappa})}+\beta\frac{I'_{\frac{t-2}{2}}(\beta\bar{\kappa})}{I_{\frac{t-2}{2}}(\beta\bar{\kappa})}+\frac{I'_{\frac{t-3}{2}}(\bar{\kappa})}{I_{\frac{t-3}{2}}(\bar{\kappa})},
\]

\[
h(\tilde{\boldsymbol{y}})=\frac{1}{\Gamma\left(\frac{t-1}{2}\right)t!2^{\frac{t-3}{2}}}\left(\frac{\beta}{\eta(\tilde{\boldsymbol{y}})}\right)^{\frac{t-2}{2}},
\]
\[
l(\bar{\kappa})=\frac{I_{\frac{t-2}{2}}(\eta(\tilde{\boldsymbol{y}})\bar{\kappa})}{I_{\frac{t-3}{2}}(\bar{\kappa})I_{\frac{t-2}{2}}(\beta\bar{\kappa})}.
\]

The detailed derivation of the predictive density of our model can
be found in Appendix B.

\section{Model Extensions}

\subsection{Incomplete rankings}

\noindent A judge may rank a set of items in accordance with some
criteria. However, in real life, some of the ranking data may be missing
either at random or by design. For example, in the former case, some
of the items may not be ranked due to the limited knowledge of the
judges. In this kind of incomplete ranking data, a missing item could
have any rank and this is called subset rankings. In another instance
called top-$k$ rankings, the judges may only rank the top 10 best
movies among several recommended. The unranked movies would in principle
receive ranks larger than $10$. In those cases, the notation $\boldsymbol{R}^{I}=(2,-,3,4,1)^{T}$
refers to a subset ranking with item 2 unranked while $\boldsymbol{R}^{I}=(2,*,*,*,1)^{T}$
represents a top two ranking with item 5 ranked first and item 1 ranked
second. 

In the usual Bayesian framework, missing data problems can be resolved
by appealing to Gibbs sampling and data augmentation methods. Let
$\left\{ \boldsymbol{R}_{1}^{I},...,\boldsymbol{R}_{N}^{I}\right\} $
be a set of $N$ observed incomplete rankings, and let $\left\{ \boldsymbol{R}_{1}^{*},...,\boldsymbol{R}_{N}^{*}\right\} $
be their unobserved complete rankings. We want to have the following
posterior distribution:
\[
p(\boldsymbol{\theta},\kappa|\boldsymbol{R}_{1}^{I},...,\boldsymbol{R}_{N}^{I})\propto p(\boldsymbol{\theta},\kappa)p(\boldsymbol{R}_{1}^{I},...,\boldsymbol{R}_{N}^{I}|\boldsymbol{\theta},\kappa),
\]
which can be achieved by Gibbs sampling based on the following two
full conditional distributions:
\[
p(\boldsymbol{R}_{1}^{*},...,\boldsymbol{R}_{N}^{*}|\boldsymbol{R}_{1}^{I},...,\boldsymbol{R}_{N}^{I},\boldsymbol{\theta},\kappa)=\prod_{i=1}^{N}p(\boldsymbol{R}_{i}^{*}|\boldsymbol{R}_{i}^{I},\boldsymbol{\theta},\kappa),
\]
\[
p(\boldsymbol{\theta},\kappa|\boldsymbol{R}_{1}^{*},...,\boldsymbol{R}_{N}^{*})\propto p(\boldsymbol{\theta},\kappa)\prod_{i=1}^{N}p(\boldsymbol{R}_{i}^{*}|\boldsymbol{\theta},\kappa).
\]

Sampling from $p(\boldsymbol{R}_{1}^{*},...,\boldsymbol{R}_{N}^{*}|\boldsymbol{R}_{1}^{I},...,\boldsymbol{R}_{N}^{I},\boldsymbol{\theta},\kappa)$
can be generated by using the Bayesian SIR method or the Bayesian
VI method which have been discussed in the previous sections.  More
concretely, we need to fill in the missing ranks for each observation
and for that we appeal to the concept of compatibility described in
\citet{alvo2014statistical} which considers for an incomplete ranking,
the class of complete order preserving rankings. For example, suppose
we observe one incomplete subset ranking $\boldsymbol{R}^{I}=(2,-,3,4,1)$.
The set of corresponding compatible rankings is $\left\{ \left(2,5,3,4,1\right)^{T},\left(2,4,3,5,1\right)^{T},\left(2,3,4,5,1\right)^{T},\left(3,2,4,5,1\right)^{T},\left(3,1,4,5,2\right)^{T}\right\} $. 

Generally speaking, let $\Omega(\boldsymbol{R}_{i}^{I})$ be the set
of complete rankings compatible with $\boldsymbol{R}_{i}^{I}$. For
an incomplete subset ranking with $k$ out of $t$ items being ranked,
we will have a total $t!/k!$ complete rankings in its compatible
set. Note that $p(\boldsymbol{R}_{i}^{*}|\boldsymbol{R}_{i}^{I},\boldsymbol{\theta},\kappa)\propto p(\boldsymbol{R}_{i}^{*}|\boldsymbol{\theta},\kappa),\:\boldsymbol{R}_{i}^{*}\in\Omega(\boldsymbol{R}_{i}^{I}).$
Obviously, direct sampling from this distribution will be tedious
for large $t$. Instead, in this paper, we use the Metropolis-Hastings
algorithm to draw samples from this distribution with the proposed
candidates generated uniformly from $\Omega(\boldsymbol{R}_{i}^{I})$.
The idea of introducing compatible rankings allows us to treat different
kinds of incomplete rankings easily. It is easy to sample uniformly
from the compatible rankings since we just need to fill-in the missing
ranks under different situations. In the case of top-$k$ rankings,
the compatibility set will be defined to ensure that the unranked
items receive rankings larger than $k$. Note that it is also possible
to use Monte Carlo EM approach to handle incomplete rankings under
a maximum likelihood setting where the Gibbs sampling is used in the
E-step (see \citet{YuLamLo2005}).

\subsection{Mixture ranking model\label{subsec:Mixture-ranking-model}}

\noindent It is quite natural to extend our simple model to that of
a mixture model in order to take into account several clusters that
may exist among heterogeneous data \citep{lee2012mixtures,kidwell2008visualizing}.
 If a population contains $G$ sub-populations (clusters), the probability
of observing a standardized ranking $\boldsymbol{y}$ under our mixture
model is given by
\[
p(\boldsymbol{y}|\boldsymbol{\kappa},\boldsymbol{\Theta},\boldsymbol{\tau})=\sum_{g=1}^{G}\tau_{g}C_{t}(\kappa_{g})\exp\left\{ \kappa_{g}\boldsymbol{\theta}_{g}^{T}\boldsymbol{y}\right\} ,
\]
where $\boldsymbol{\tau}=(\tau_{1},\ldots,\tau_{G})$, with $\tau_{g}$
representing the proportion or the mixture weights for the $g$th
sub-population whereas $\boldsymbol{\Theta}=\left(\boldsymbol{\theta}_{1},\ldots,\boldsymbol{\theta}_{G}\right)$
and $\boldsymbol{\kappa}=\left(\kappa_{1},\ldots,\kappa_{G}\right)$,
with $\boldsymbol{\theta}_{g}$ and $\kappa_{g}$ are the directional
and concentration parameters in the $g$th sub-population respectively.
To obtain the MLE of this mixture model, we may extend the approach
described in Section \ref{subsec:MLE-of-osimple-model} using the
traditional EM algorithm. 

The variational inference approach for this mixture model follows
the method of \citet{taghia2014bayesian}. Given a random sample of
$N$ complete standardized rankings $\boldsymbol{Y}=\left\{ \boldsymbol{y}_{1},\ldots,\boldsymbol{y}_{N}\right\} $
drawn from $p(\boldsymbol{y}|\boldsymbol{\kappa},\boldsymbol{\Theta},\boldsymbol{\tau})$.
We first introduce a set of binary latent variables $\boldsymbol{Z}=\left\{ z_{ig}\right\} $
where $i=1,\ldots,N$, $g=1,\ldots,,G$ where $z_{ig}=1$ indicates
the observed ranking $\boldsymbol{y}_{i}$ belongs to the $g$th sub-population.
Thus the generative model may be written as 
\[
p(\boldsymbol{Y},\boldsymbol{Z},\boldsymbol{\tau},\boldsymbol{\Theta},\boldsymbol{\kappa})=p(\boldsymbol{Y}|\boldsymbol{Z},\boldsymbol{\Theta},\boldsymbol{\kappa})p(\boldsymbol{\Theta},\boldsymbol{\kappa})p(\boldsymbol{Z}|\boldsymbol{\tau})p(\boldsymbol{\tau}),
\]
where
\begin{eqnarray*}
p(\boldsymbol{Y}|\boldsymbol{Z},\boldsymbol{\Theta},\boldsymbol{\kappa}) & = & \prod_{i=1}^{N}\prod_{g=1}^{G}\left(C_{t}(\kappa_{g})\exp\left\{ \kappa_{g}\boldsymbol{\theta}_{g}^{T}\boldsymbol{y}_{i}\right\} \right)^{z_{ig}}\\
p(\boldsymbol{Z}|\boldsymbol{\tau}) & = & \prod_{i=1}^{N}\prod_{g=1}^{G}\tau_{g}^{z_{ig}}
\end{eqnarray*}

A Dirichlet distribution with prior vector parameters $d_{0,g}$ is
considered for the prior distribution of $\boldsymbol{\tau}$ :
\[
p(\boldsymbol{\tau})=\frac{\Gamma(\sum_{g=1}^{G}d_{0,g})}{\prod_{g=1}^{G}\varGamma\left(d_{0,g}\right)}\prod_{g=1}^{G}\tau_{k}^{d_{o,g}-1}.
\]
The prior distribution for $\left(\boldsymbol{\Theta},\boldsymbol{\kappa}\right)$
is the conditional von Mises-Fisher distribution for $\boldsymbol{\Theta}|\boldsymbol{\kappa}$
and the marginal Gamma distribution for $\boldsymbol{\kappa}$:
\[
p(\boldsymbol{\Theta},\boldsymbol{\kappa})=\prod_{g=1}^{G}vMF(\boldsymbol{\theta}|\boldsymbol{m}_{0,g},\beta_{0,g}\kappa_{g})Gamma(\kappa_{g}|a_{0,g},b_{0,g}),
\]
where $\boldsymbol{m}_{0,g},\beta_{0,g},a_{0,g},b_{0,g}$ are the
prior parameters of the $g$th sub-population. Using the similar technique
in Section \ref{subsec:Optimization-of-the-VI} to optimize the evidence
lower bound given by
\begin{equation}
\mathcal{L_{M}}(q)=E_{q(\boldsymbol{Z},\boldsymbol{\Theta},\boldsymbol{\kappa},\boldsymbol{\tau})}\left[\ln\frac{p(\boldsymbol{Y}|\boldsymbol{Z},\boldsymbol{\Theta},\boldsymbol{\kappa})p(\boldsymbol{\Theta},\boldsymbol{\kappa})p(\boldsymbol{Z}|\boldsymbol{\tau})p(\boldsymbol{\tau})}{q(\boldsymbol{Z})q(\boldsymbol{\Theta}|\boldsymbol{\kappa})q(\boldsymbol{\kappa})q(\boldsymbol{\tau})}\right],\label{eq:evidence_lower_bound-mixture-1}
\end{equation}
we can derive the optimal posterior distribution of each parameter. 

It is not difficult to see that the optimal posterior distribution
for $q(\boldsymbol{\tau})$ is recognized to be a Dirichlet distribution
with parameter $d_{g}$:
\begin{equation}
d_{g}=d_{0,g}+\sum_{i=1}^{N}p_{ig},\label{eq:update_d_g}
\end{equation}
where
\begin{equation}
p_{ig}=\frac{\exp(\rho_{ig})}{\sum_{j=1}^{G}\exp(\rho_{ij})},\label{eq:update_p_ig}
\end{equation}
\begin{align}
\rho_{ig}=\frac{t-3}{2}E_{q(\kappa)}(\ln\kappa_{g})+E_{q(\boldsymbol{\tau})}\left(\ln\tau_{g}\right)+E_{q(\boldsymbol{\Theta},\boldsymbol{\kappa})}\left(\kappa_{g}\boldsymbol{\theta}_{g}^{T}\boldsymbol{y}_{i}\right)-\ln\left[2^{\frac{t-3}{2}}t!\Gamma\left(\frac{t-1}{2}\right)\right]\label{eq:update_rho_ig}\\
-\ln I_{\frac{t-3}{2}}\left(\bar{\kappa}_{g}\right)-\left(\frac{\partial}{\partial\kappa_{g}}\ln I_{\frac{t-3}{2}}\left(\bar{\kappa}_{g}\right)\right)\left[E_{q(\boldsymbol{\kappa})}\kappa_{g}-\bar{\kappa}_{g}\right],\nonumber 
\end{align}
\begin{equation}
\bar{\kappa}_{g}=\begin{cases}
\frac{a_{g}-1}{b_{g}} & \mbox{if }a_{g}>1\\
\frac{a_{g}}{b_{g}} & \mbox{otherwise}
\end{cases},\label{eq:kappa bar_g}
\end{equation}
and the optimal posterior distribution of $q(\boldsymbol{\Theta}|\boldsymbol{\kappa})$
can be written as von Mises-Fisher distribution:
\[
q^{*}(\boldsymbol{\Theta}|\boldsymbol{\kappa})=\prod_{g=1}^{G}vMF(\boldsymbol{\theta}_{g}|\boldsymbol{m}_{g},\kappa_{g}\beta_{g})
\]

\begin{equation}
\beta_{g}=\left\Vert \beta_{0,g}\boldsymbol{m}_{0,g}+\sum_{i=1}^{N}p_{ig}\boldsymbol{y}_{i}\right\Vert ,\label{eq:Update_beta_g}
\end{equation}
\begin{equation}
\boldsymbol{m}_{g}=\left(\beta_{0,g}\boldsymbol{m}_{0,g}+\sum_{i=1}^{N}p_{ig}\boldsymbol{y}_{i}\right)\beta_{g}^{-1}.\label{eq:Update_m_g}
\end{equation}
Also, the optimal distribution of $q^{*}(\boldsymbol{\kappa})=\prod_{g=1}^{G}q^{*}(\kappa_{g})$
can be recognized as independent Gamma distributions:

\[
q^{*}(\kappa_{g})=Gamma(\kappa_{g}|a_{g},b_{g}),
\]

\begin{equation}
a_{g}=a_{0,g}+\left(\frac{t-3}{2}\right)\sum_{i=1}^{N}p_{ig}+\beta_{g}\bar{\kappa}_{g}\left[\frac{\partial}{\partial\beta_{g}\kappa_{g}}\ln I_{\frac{t-2}{2}}(\beta_{g}\bar{\kappa}_{g})\right],\label{eq:update_postrerior_a-Mixture}
\end{equation}
\begin{equation}
b_{g}=b_{0,g}+\left(\sum_{i=1}^{N}p_{ig}\right)\frac{\partial}{\partial\kappa_{g}}\ln I_{\frac{t-3}{2}}(\bar{\kappa}_{g})+\beta_{0,g}\left[\frac{\partial}{\partial\beta_{0,g}\kappa_{g}}\ln I_{\frac{t-2}{2}}(\beta_{0,g}\bar{\kappa}_{g})\right],\label{eq:Update_posterior_b-mixture}
\end{equation}
and finally the optimal variational posterior distribution for $\boldsymbol{Z}$
is recognized as a multinomial distribution:
\[
q^{*}(\boldsymbol{Z})=\prod_{i=1}^{N}\prod_{g=1}^{G}p_{ig}^{z_{ig}}.
\]

The detailed derivation of the optimization of the mixture model can
be found in Appendix C. A summary of the algorithm for estimation
for this mixture model  is shown in Algorithm \ref{alg:Bayesian-Estimation-our-mixture-model}.

\begin{algorithm}[h]
\textbf{\small{}Input:}{\small{} Scaled $\boldsymbol{Y}=\left\{ \boldsymbol{y}_{1},...,\boldsymbol{y}_{n}\right\} $}{\small \par}

\textbf{\small{}Step 1: Initialization}{\small \par}
\begin{enumerate}
\item {\small{}Set the prior parameters: $d_{0,g}$, $\beta_{0,g}$, $\boldsymbol{m}_{0,g}$,
$a_{0,g}$ , $b_{0,g}$ and number of clusters $G$.}{\small \par}
\item {\small{}Initialize $p_{ig}=\frac{1}{G}$ and the initial value of
$\bar{\kappa}_{g}=\frac{a_{0,g}}{b_{0,g}}.$}{\small \par}
\end{enumerate}
\textbf{\small{}Step 2: Optimization of the posterior distribution}{\small \par}

\textbf{\small{}repeat}{\small \par}
\begin{enumerate}
\item {\small{}Update posterior parameters $d_{g}$, $\beta_{g}$, $\boldsymbol{m}_{g}$,
$a_{g}$ , $b_{g}$ by (\ref{eq:update_d_g}), (\ref{eq:Update_beta_g}),
(\ref{eq:Update_m_g}), (\ref{eq:update_postrerior_a-Mixture}) and
(\ref{eq:Update_posterior_b-mixture}).}{\small \par}
\item {\small{}Update $p_{ig}$ by (\ref{eq:update_p_ig}) and (\ref{eq:update_rho_ig}).}{\small \par}
\item {\small{}Update $\bar{\kappa}_{g}$ by (\ref{eq:kappa bar_g}).}{\small \par}
\end{enumerate}
\textbf{\small{}until convergence}{\small \par}

\caption{\label{alg:Bayesian-Estimation-our-mixture-model}Bayesian Estimation
using variational inference of our mixture ranking model.}
\end{algorithm}

\section{Simulation Studies}

\subsection{Comparison of the posterior distributions obtained by Bayesian SIR
method and variational inference approach}

\begin{figure}[h]
\begin{centering}
\includegraphics[scale=0.5]{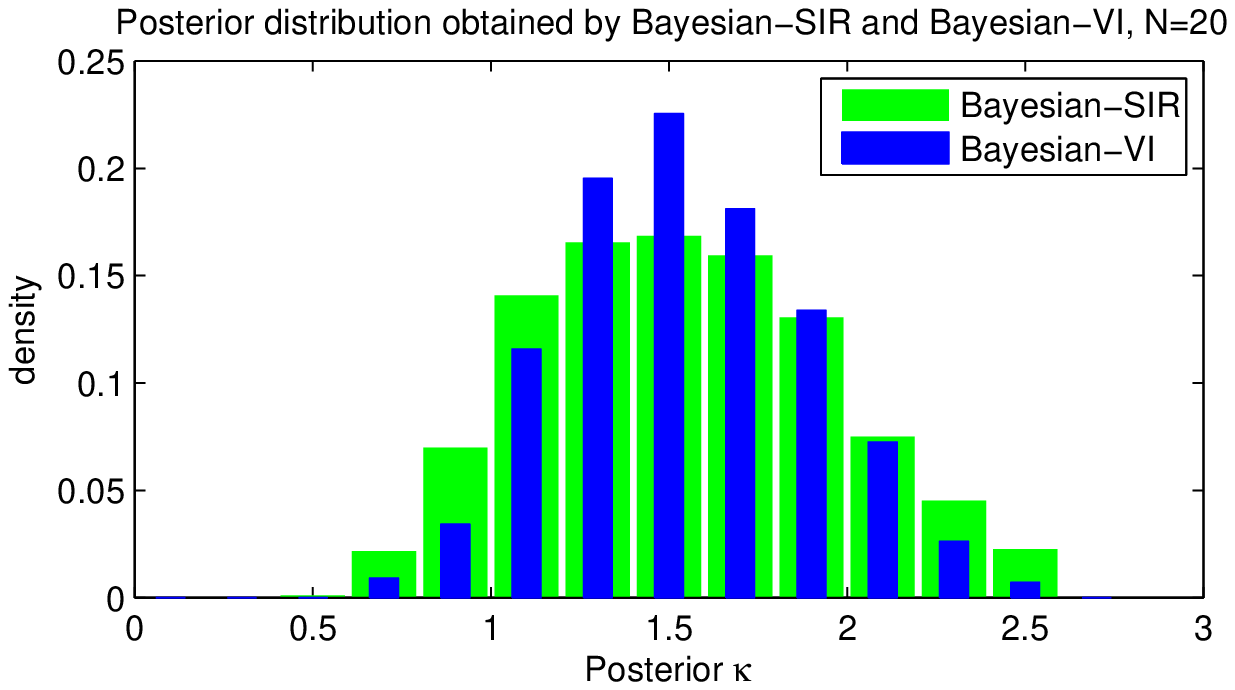}\includegraphics[scale=0.5]{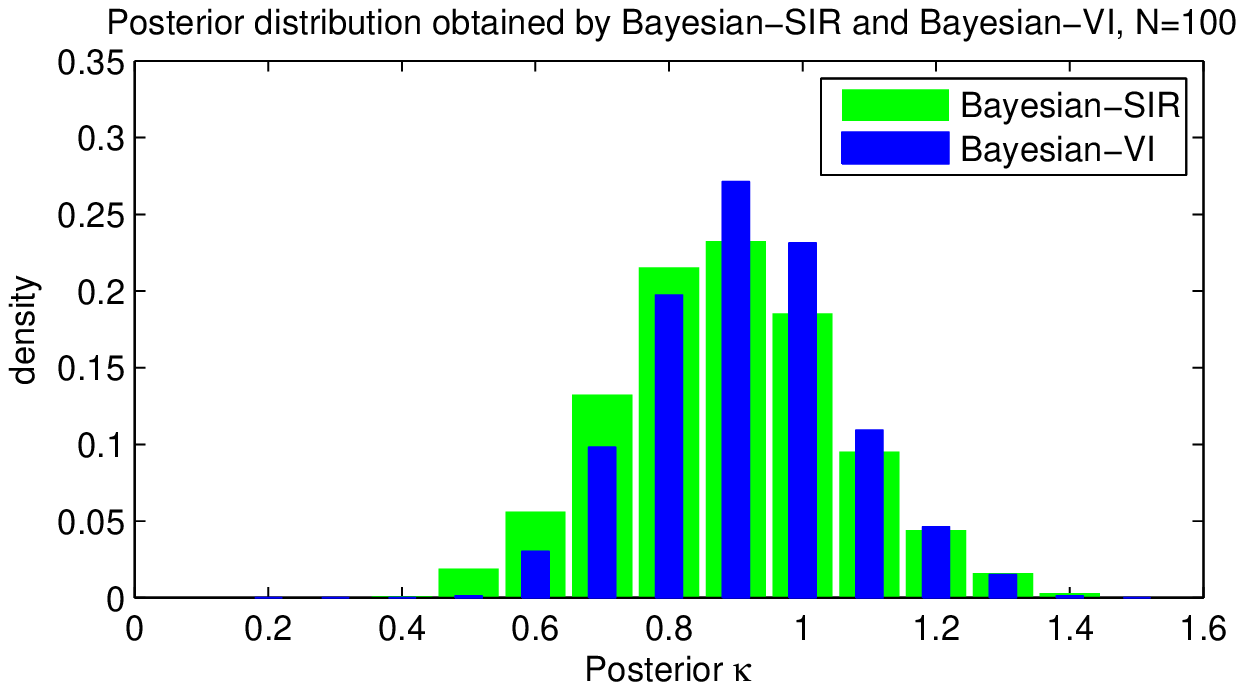}
\par\end{centering}
\begin{centering}
\includegraphics[scale=0.5]{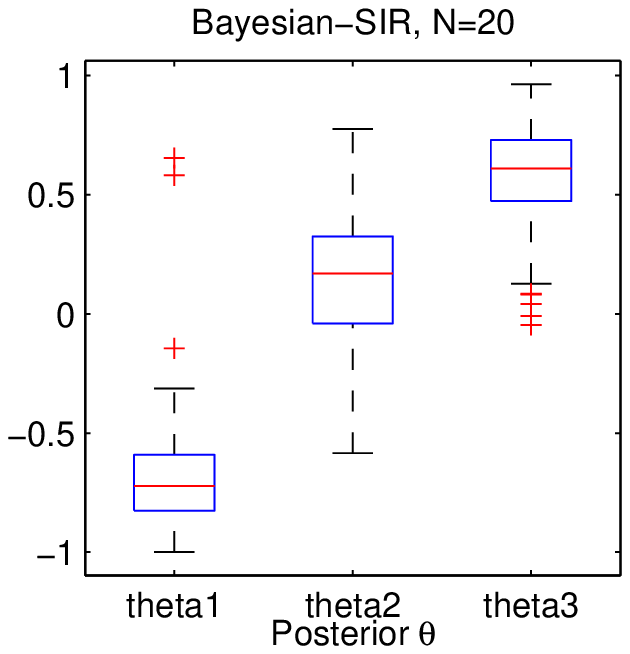}\includegraphics[scale=0.5]{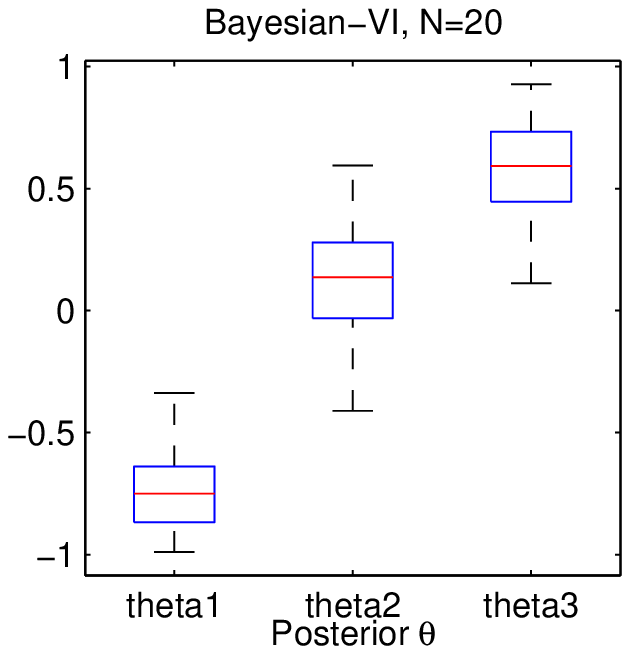}\includegraphics[scale=0.5]{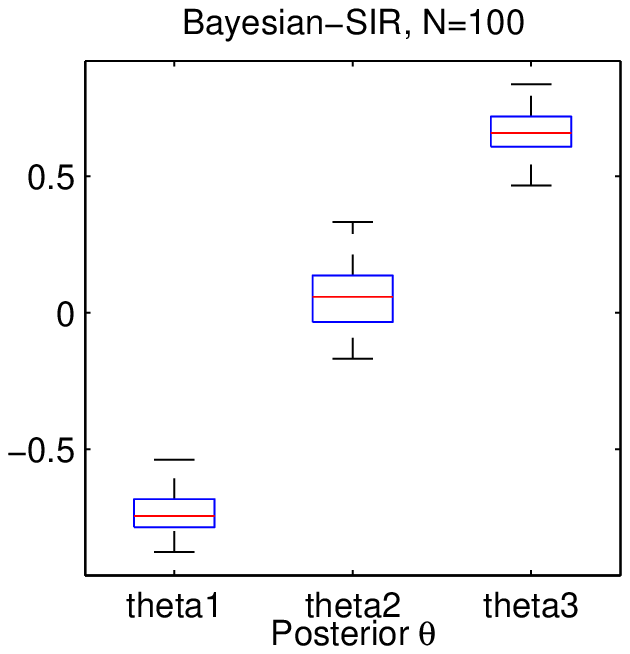}\includegraphics[scale=0.5]{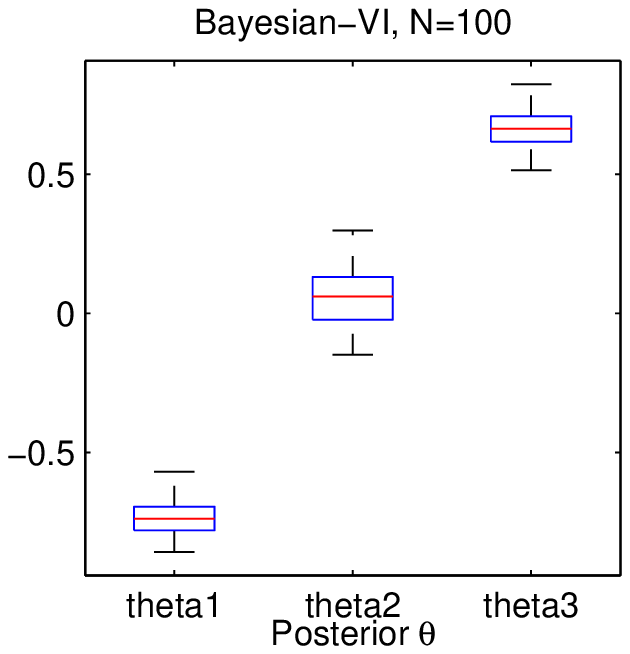}
\par\end{centering}
\caption{\label{fig:Comparison-posterior_distribution_SIR_VI}Comparison of
the posterior distribution obtained by Bayesian SIR method and the
approximate posterior distribution by variational inference approach.
The comparison is illustrated for different data sizes of $N=20$
(left) and $N=100$ (right).}
\end{figure}

\noindent Since we use a factorized approximation for the posterior
distribution in the variational inference approach, it is of interest
to compare the true posterior distribution with its approximation
obtained using the variational inference approach. We simulated two
data sets with $\kappa=1$, $\boldsymbol{\theta}=\left(-0.71,0,0.71\right)^{T},$
$t=3$ and different data sizes of $N=20,100.$ We generated samples
from the posterior distribution by SIR method in Section \ref{subsec:Bayesian-method-SIR}
using a gamma density with mean $\hat{\kappa}_{MLE}$ and variance
equal to $0.2$ as the proposal density. We then applied the variational
approach in Algorithm \ref{alg:Bayesian-Estimation-our_model} and
generated samples from the corresponding posterior distribution. Figure
\ref{fig:Comparison-posterior_distribution_SIR_VI} exhibits the histogram
and box-plot for the posterior distribution of $\kappa$ and $\boldsymbol{\theta}$.

From Figure \ref{fig:Comparison-posterior_distribution_SIR_VI}, we
see that the posterior distribution using the Bayesian-VI is very
close to the posterior distribution obtained by the Bayesian-SIR method.
When the sample size is small ($N=20$), there are more outliers for
the Bayesian-SIR method while the posterior $\kappa$ for the Bayesian-VI
method seems to be more concentrated. When the sample size is large,
the posterior estimates of $\boldsymbol{\theta}$ and $\kappa$ become
more accurate and Bayesian-VI is closer to the posterior distribution
obtained by the Bayesian-SIR method. We calculate the symmetric variant
of the Kullback-Leibler divergence (KLD) between two distributions
obtained by two methods for posterior $\kappa$. The symmetric KLD
are 0.45 for the $N=20$ case and 0.44 for the $N=100$ case. More
simulations for different settings of parameters can be found in Appendix
D.

\subsection{Experiments with different sample sizes\label{subsec:Experiments-different_N}}

\noindent We also evaluated the performance of the three estimating
algorithms for our model when the sample size $N$ is allowed to vary
from 25 to 500. We simulated three different data sets with the number
of items being ranked $t=10,20,50.$ The true $\boldsymbol{\theta}$
is a random unit vector. Since our model is not a standard distribution,
we use the random-walk Metropolis algorithm to draw samples from it
\citep{liu2008monte}.

We compared the performance of the MLE method, the Bayesian method
with SIR for posterior sampling (Bayesian-SIR) and the Bayesian VI.
We chose non-informative priors for both Bayesian-SIR and Bayesian-VI.
Specifically, the prior parameter $\boldsymbol{m}_{0}$ is chosen
uniformly whereas $\beta_{0}$, $a_{0}$ and $b_{0}$ are chosen to
be small numbers close to zero. For the MLE method, we perform Newton-Raphson
iterations to get a more accurate $\kappa$. For the posterior distribution
of $\kappa$ in the Bayesian-SIR method. we used a Gamma density with
mean $\hat{\kappa}_{MLE}$ and variance 1 as the proposal density
to sample $1,000$ observations of $\kappa$ from $10,000$ candidates. 

\begin{figure}
\begin{centering}
\includegraphics[height=4cm]{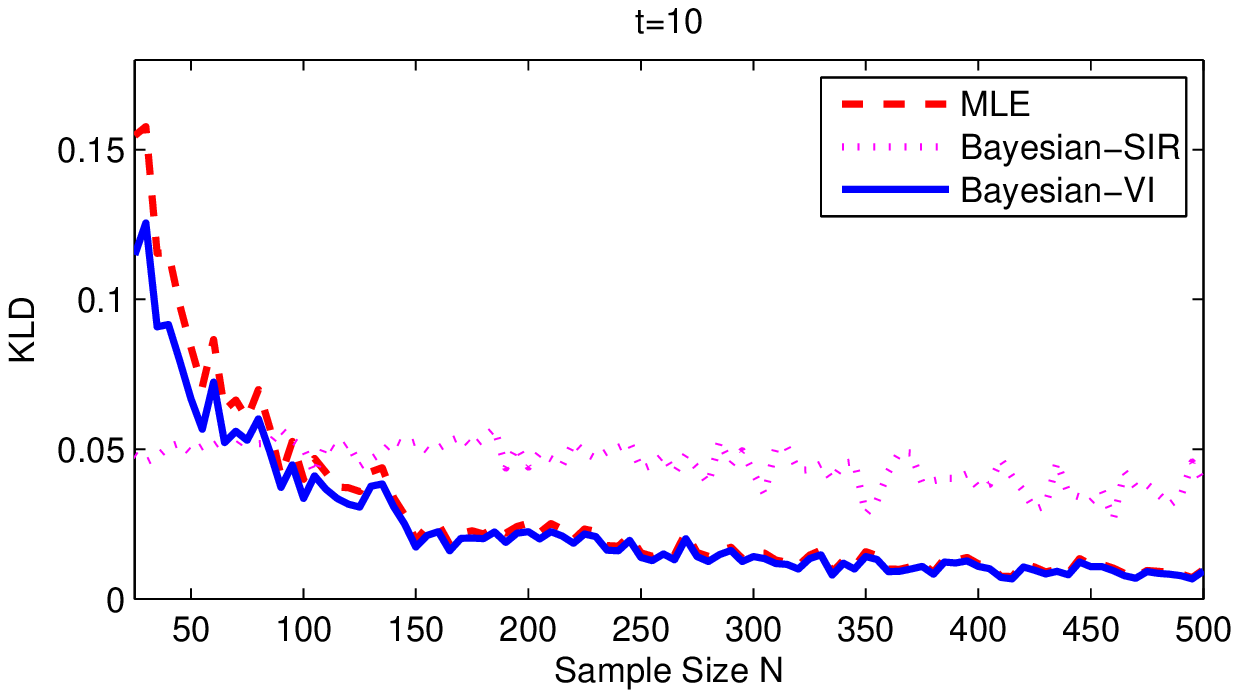}
\par\end{centering}
\begin{centering}
\includegraphics[height=4cm]{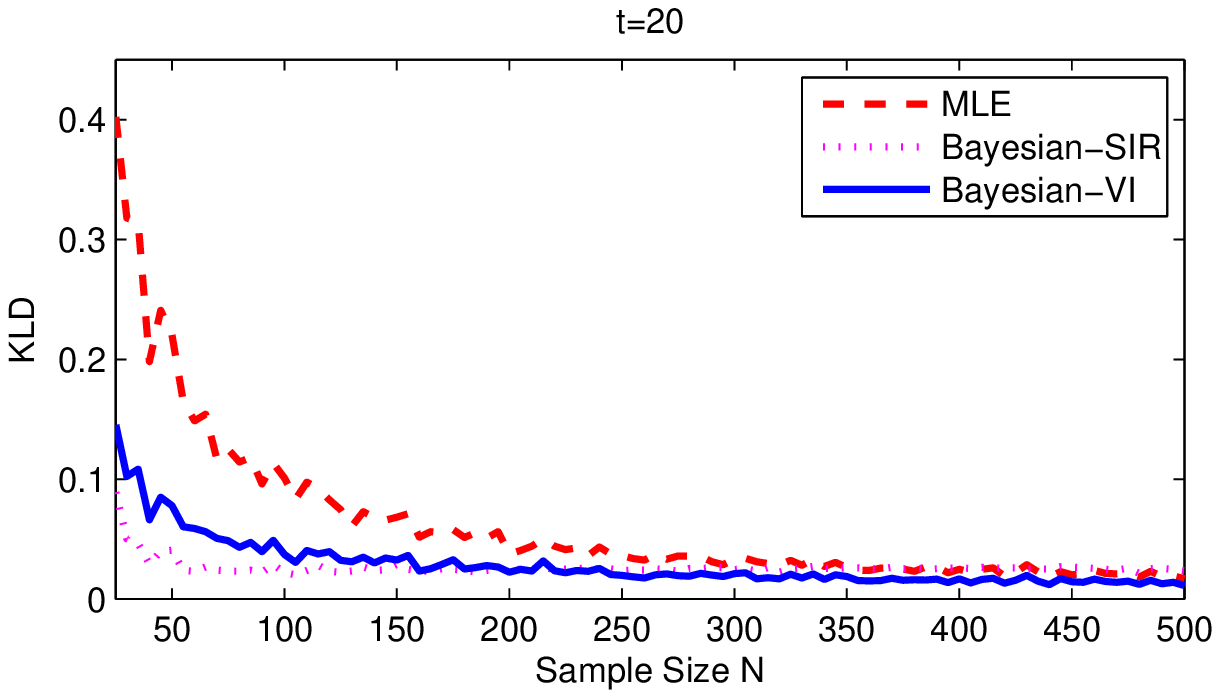}
\par\end{centering}
\begin{centering}
\includegraphics[height=4cm]{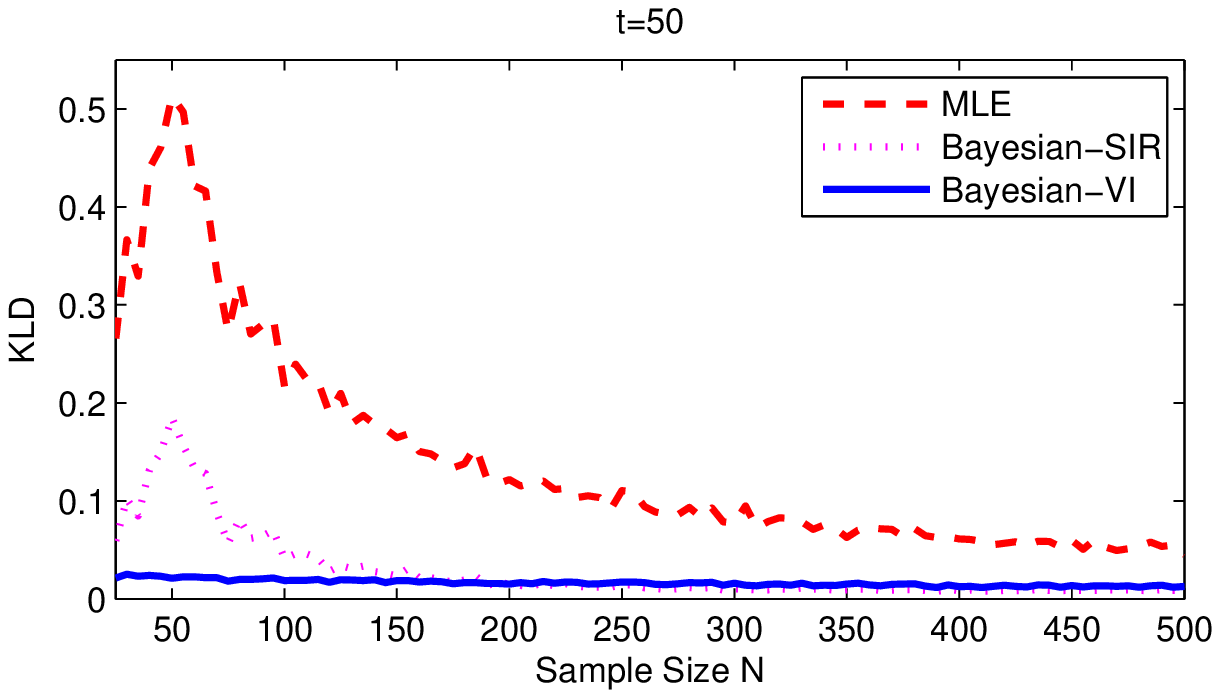}
\par\end{centering}
\caption{\label{fig:fig:KLD-Simple_model_diff_N}The KLD values between true
model and its estimated model versus the Sample size $N$ for $t=10$
(top), $t=20$ (middle) and $t=50$ (bottom). The estimated model
is given by the MLE method, Bayesian method with SIR for posterior
sampling (Bayesian-SIR) and the Bayesian method with variational approach
(Bayesian-VI). Each experiment is repeated 10 times and the average
results are shown. }
\end{figure}

We calculated the Kullback-Leibler divergence (KLD) of the true model
from the estimated model, in which the model parameters are the point
estimates derived by either MLE or the posterior mean of the Bayesian
method. A smaller value of KLD implies higher accuracy of the estimation
method. Each experiment is repeated 10 times to smooth out the effect
of random initialization and the average results are shown in Figure
\ref{fig:fig:KLD-Simple_model_diff_N}. 

It can be seen from Figure \ref{fig:fig:KLD-Simple_model_diff_N}
that large sample sizes lead to lower KLD values for the three algorithms
as expected. From the comparison, when $t$ is small $(t=10)$, the
Bayesian method with variational approach (Bayesian-VI) performs similar
to the MLE method and works better than Bayesian-SIR. The failure
of Bayesian-SIR may be the result of the variance of the proposal
gamma density being too large compared to the variance of the true
posterior distribution of $\kappa.$ Thus this improper choice of
variance will slow down the convergence of the MCMC model. When $t$
is large ($t=20$ \& $50$), Bayesian-VI and Bayesian-SIR are very
close while the MLE method doesn't work well when the sample size
is small. When $N$ is large, the three approaches tend to converge
to fairly similar results. Even for $t=50$ and $N=500$, the MLE
method still performs slightly poorer than the Bayesian counterparts.
As a whole. the Bayesian-VI generally performs the best for different
sets of $t$ and $N$.

\begin{figure}
\begin{centering}
\includegraphics[height=4cm]{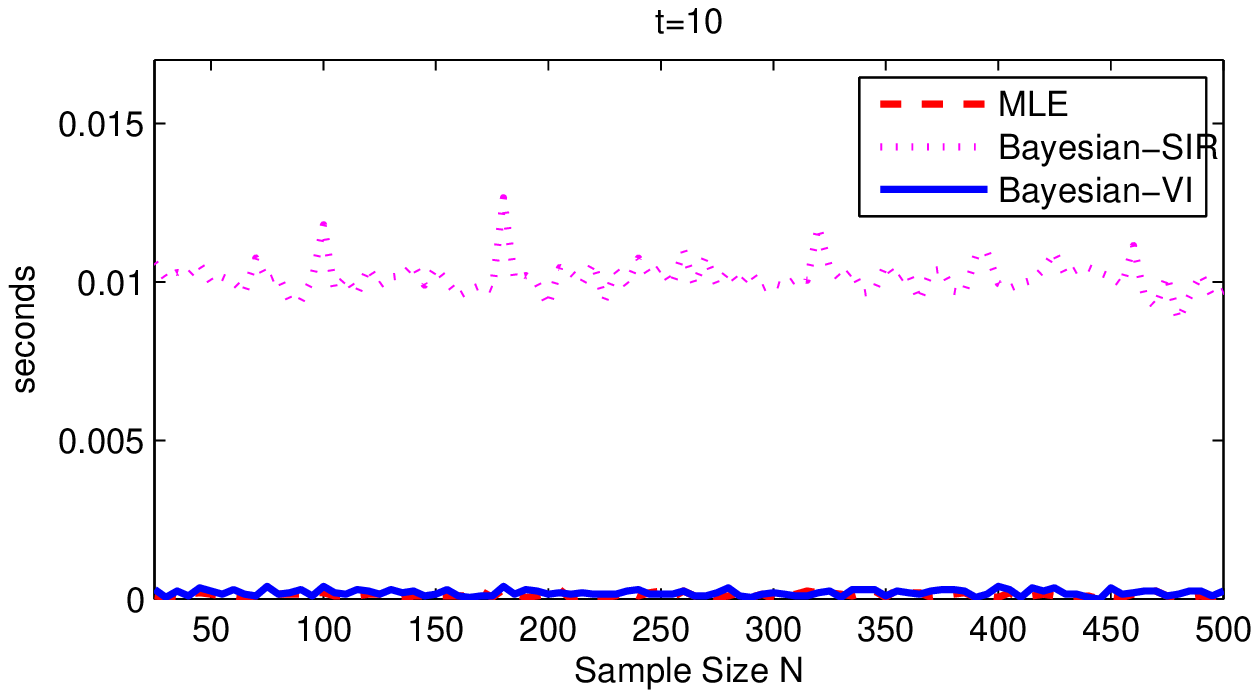}\includegraphics[height=4cm]{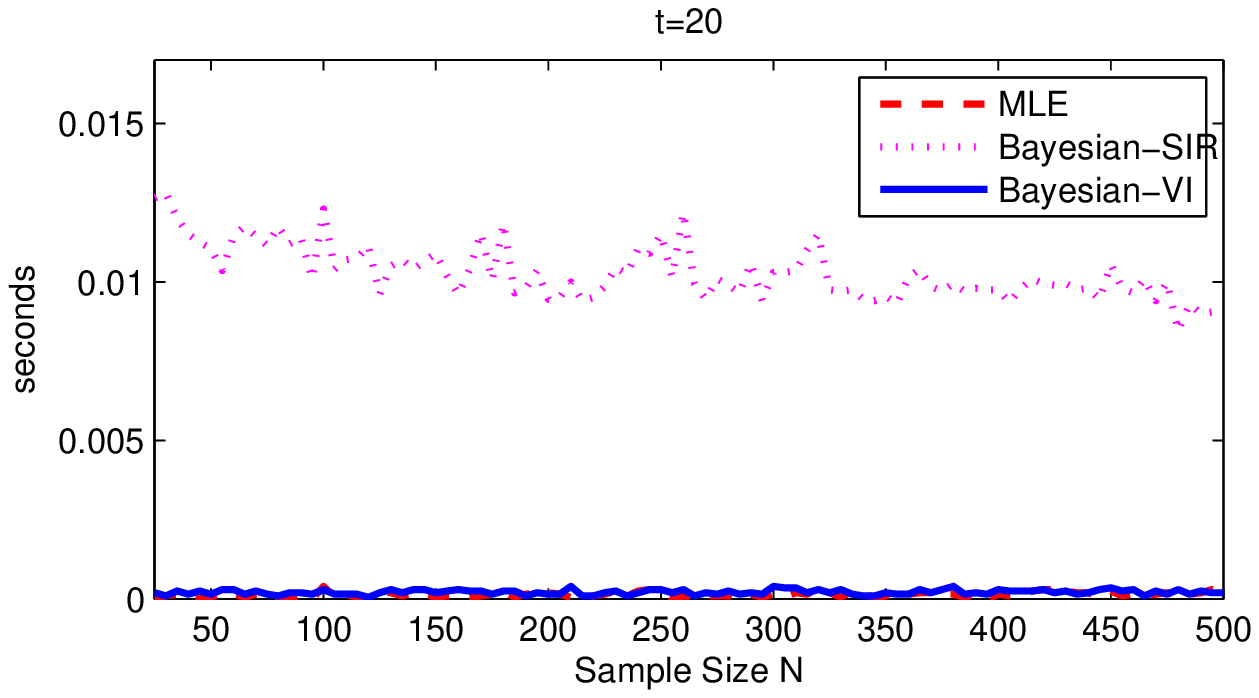}
\par\end{centering}
\caption{\label{fig:Average-Computation-times_diff_N}Average computation times
(seconds) for one run versus the Sample size $N$ by the MLE method,
Bayesian method with SIR (Bayesian-SIR) and the Bayesian method with
variational approach (Bayesian-VI). All the simulations were conducted
on a PC with 4.0 GHz quad-core CPU.}
\end{figure}

We also computed the average computation time for each set of experiments.
From Figure \ref{fig:Average-Computation-times_diff_N}, we see that
the computation time for Bayesian-SIR is the slowest as expected since
it is an MCMC sampling method. The speeds of Bayesian-VI and MLE are
quite similar and they are about 50 to 100 times faster than Bayesian-SIR.
All the simulations were conducted on a PC with 4.0 GHz quad-core
CPU.

\subsection{Experiments with different data dimensions \label{subsec:Effect_t_simple_model}}

\begin{figure}
\begin{centering}
\includegraphics[height=4cm]{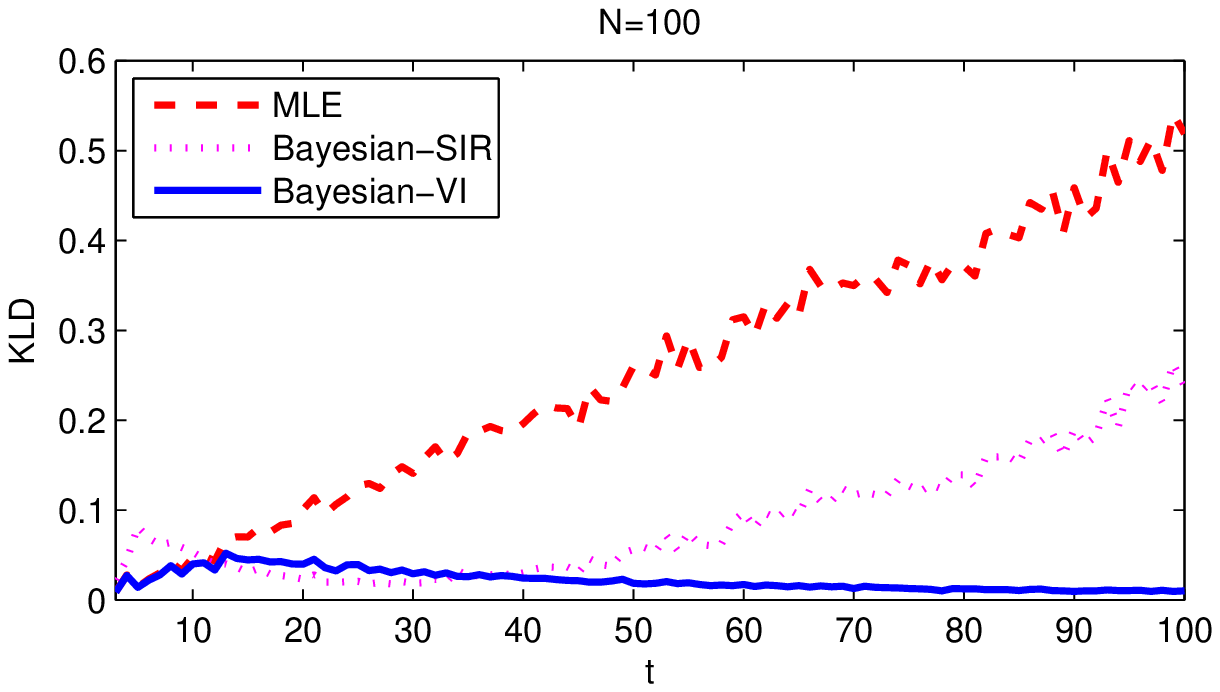}
\par\end{centering}
\begin{centering}
\includegraphics[height=4cm]{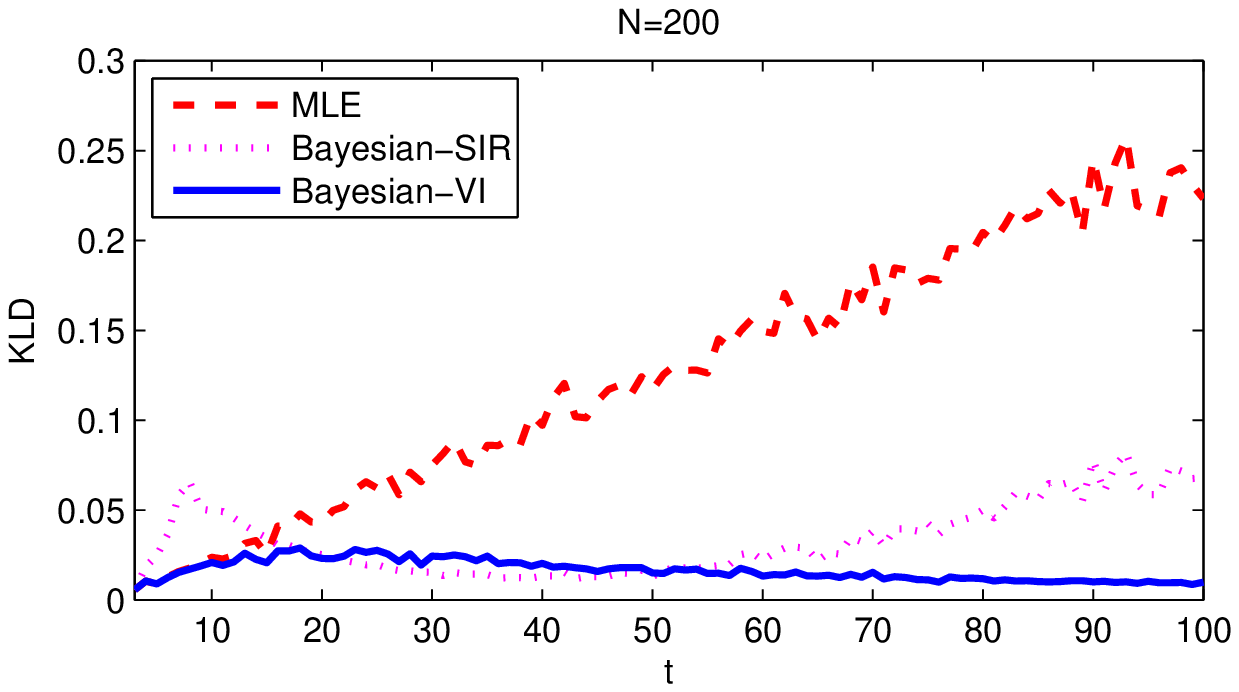}
\par\end{centering}
\begin{centering}
\includegraphics[height=4cm]{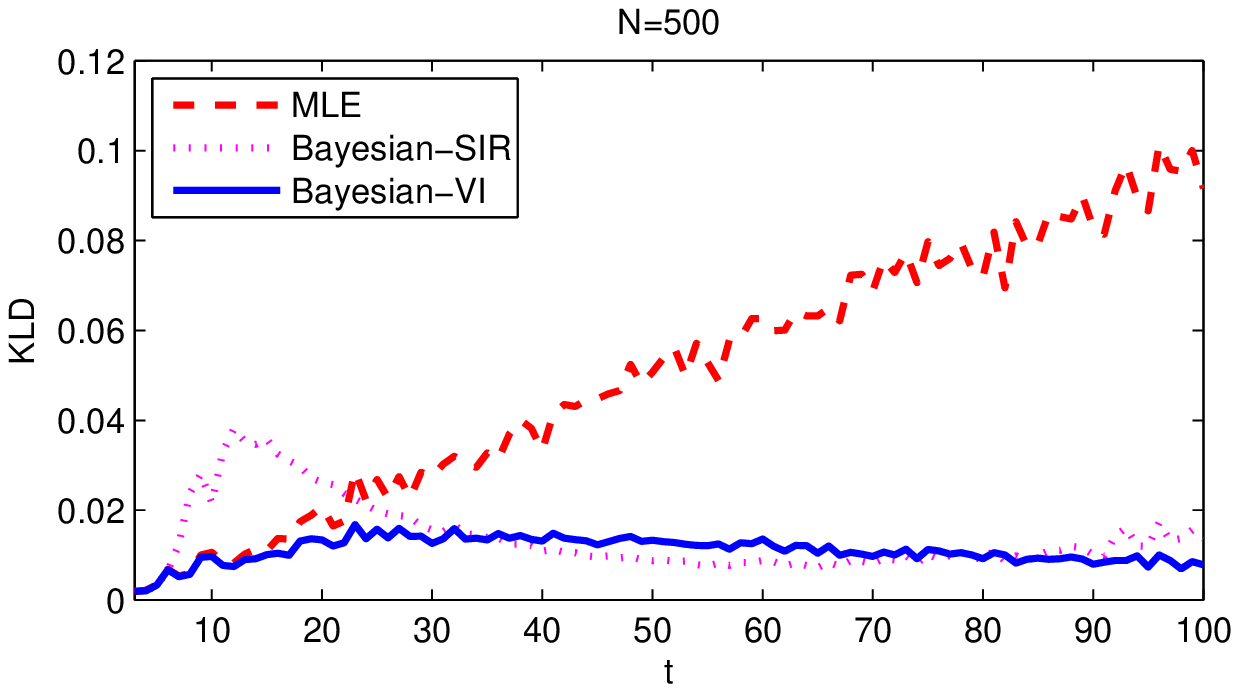}
\par\end{centering}
\caption{\label{fig:KLD-Simple_model_diff_t}The KLD values between true model
and its estimated model versus the data dimension $t$ (number of
items being ranked) for data sizes of $N=100$ (top), $N=200$ (middle)
and $N=500$ (bottom). The estimated model is given by the MLE method
(MLE), Bayesian method with SIR for posterior sampling (Bayes-SIR)
and the Bayesian method with variational approach (Bayes-VI). Each
experiment is repeated 10 times and the average results are shown. }
\end{figure}

\noindent In the following experiments we compared the performances
of the different approaches as the number $t$ of items ranked varies
from 3 to 100. We set $\kappa=1$ and chose the true $\boldsymbol{\theta}$
to be a random unit vector. We chose $N=100,200,500$. The detailed
simulation settings are the same as Section \ref{subsec:Experiments-different_N}.
For evaluation, we again calculated the Kullback-Leibler divergence
(KLD) of the true model from the estimated model shown in Figure \ref{fig:KLD-Simple_model_diff_t}.
Each experiment is repeated 10 times and the average results are shown
to smooth out the effect of random initialization. 

It is seen that large values of $t$ lead to the failure of the MLE
method since there will be more parameters than data-points. The Bayesian-SIR
method also encounters problems when $t$ is large compared to $N.$
This may be because the selected proposal density (variance $=1$)
in the SIR method may be inappropriate when $t$ is large. From the
comparison, the Bayesian -VI method has lower KLD values for large
$t$. When $t$ is small, the MLE and Bayesian-VI have similar results
while the Bayesian-SIR method has higher KLD.

\begin{figure}
\begin{centering}
\includegraphics[height=4cm]{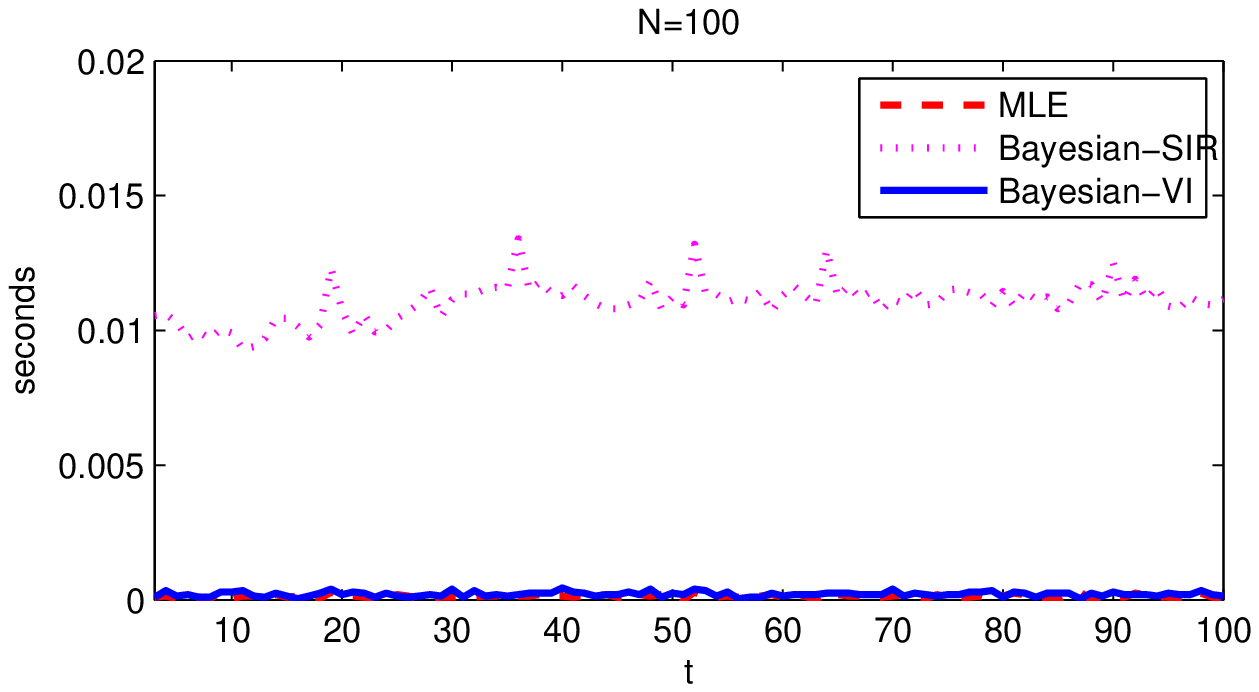}\includegraphics[height=4cm]{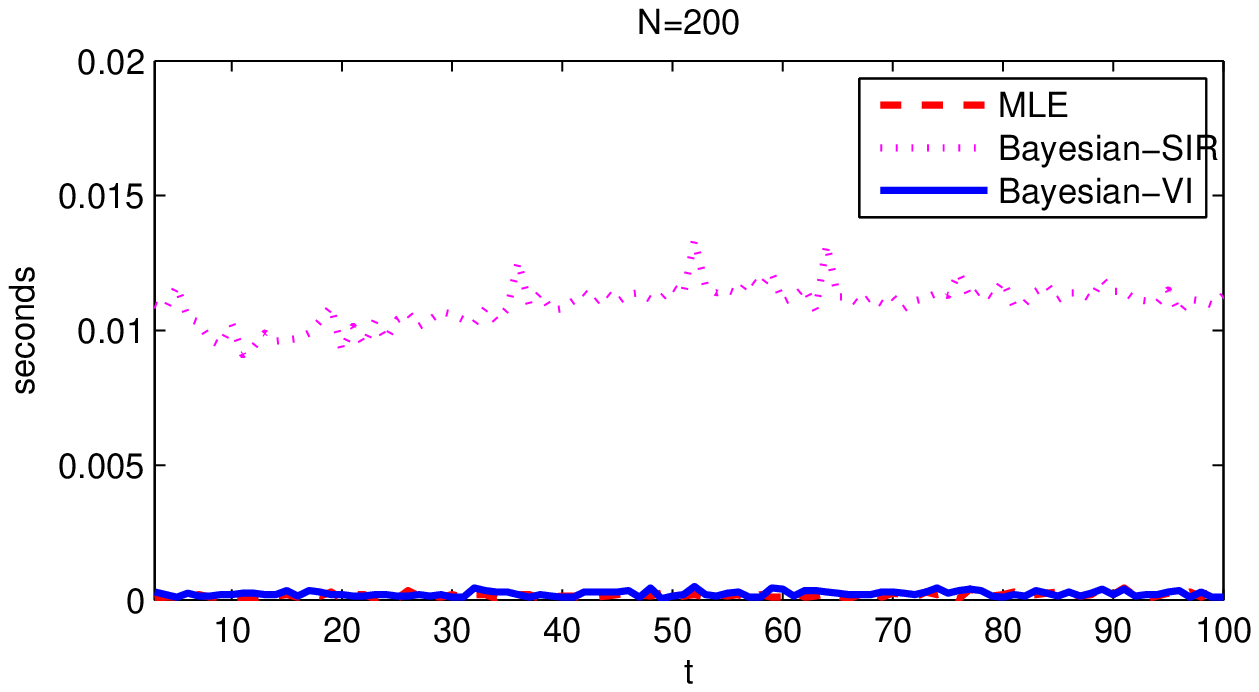}
\par\end{centering}
\caption{\label{fig:Average-Computation-times_diff_t}Average computation times
(seconds) for one run versus the data dimension $t$ by the MLE approach
(MLE), Bayesian method with SIR (Bayesian-SIR) and the Bayesian method
with variational approach (Bayesian-VI). All the simulations were
conducted on a PC with 4.0 GHz quad-core CPU. }
\end{figure}

We also computed the average computation time for each set of experiments.
From Figure \ref{fig:Average-Computation-times_diff_t}, the speed
of convergence for Bayesian-VI and MLE are quite similar and they
are about 50 to 100 times faster than Bayesian-SIR. All the simulations
are conducted on a PC with 4 core 4.0 GHz CPU.

\subsection{Simulation for the estimation of the predictive density}

\begin{figure}[h]
\begin{centering}
\includegraphics[height=4cm]{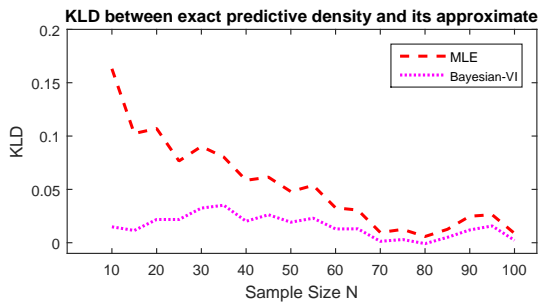}
\par\end{centering}
\caption{\label{fig:The-KLD-predictive_density}The KLD between the exact predictive
density $p(\tilde{y}|Y)$ and the approximate predictive densities
by Bayesian method with variational approach (Bayesian-VI) and MLE
method (MLE). Each experiment is repeated 10 times and the average
results are shown. }
\end{figure}

\noindent In this experiment, we compared the accuracy of the approximated
predictive density between the Bayesian-VI and MLE methods. We simulated
data from our model with $\kappa=1$ and $t=5$. The true $\boldsymbol{\theta}$
is a random unit vector. For each set of simulation, we considered
sample sizes ranging from 10 to 100. We also calculated the Kullback-Leibler
divergence (KLD) from the true posterior distribution to the approximate
predictive density. The true posterior predictive distribution is
calculated by Monte Carlo integration (\ref{eq:exact_predictive_density-1})
numerically from the true posterior distribution obtained by the SIR
method with size $10^{5}$. We calculated the approximated predictive
density using the MLE method. Each experiment was repeated 10 times
and the average results are shown in Figure \ref{fig:The-KLD-predictive_density}.

From Figure \ref{fig:The-KLD-predictive_density}, we see that the
KLD for both methods decreases with increasing sample size as expected.
However, for small training data, Bayesian-VI performs much better
than the MLE method.

\subsection{Simulation for incomplete rankings}

\begin{figure}[h]
\begin{centering}
\includegraphics[height=4cm]{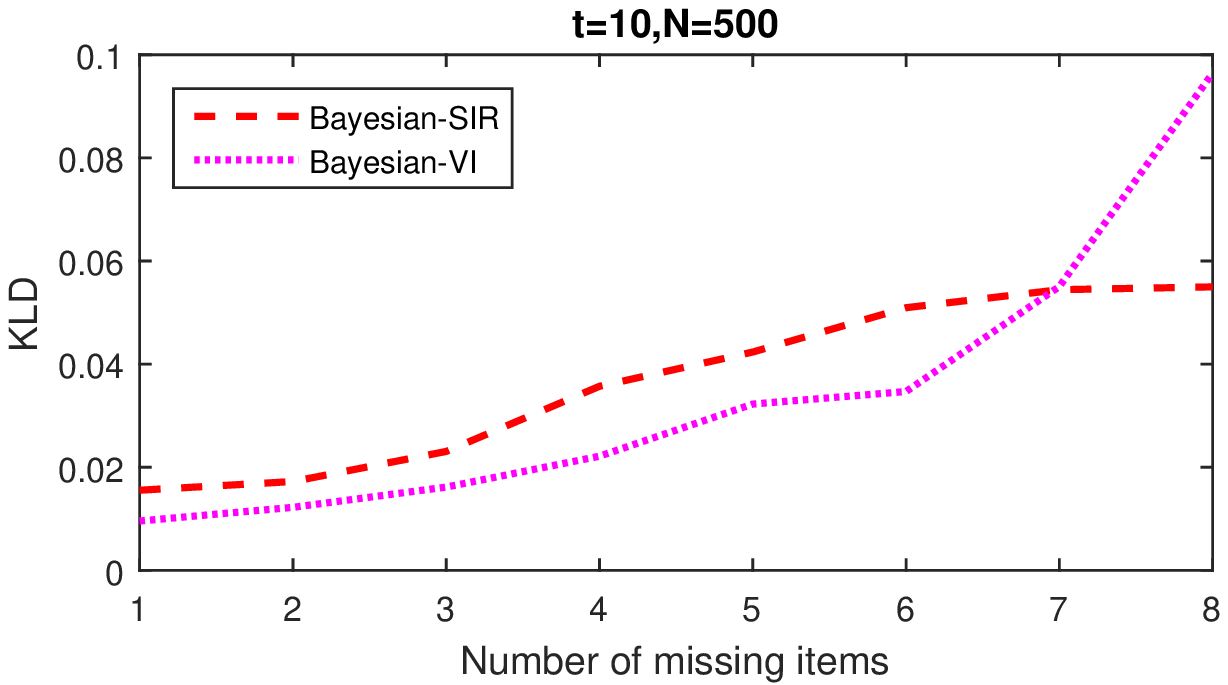}
\par\end{centering}
\begin{centering}
\includegraphics[height=4cm]{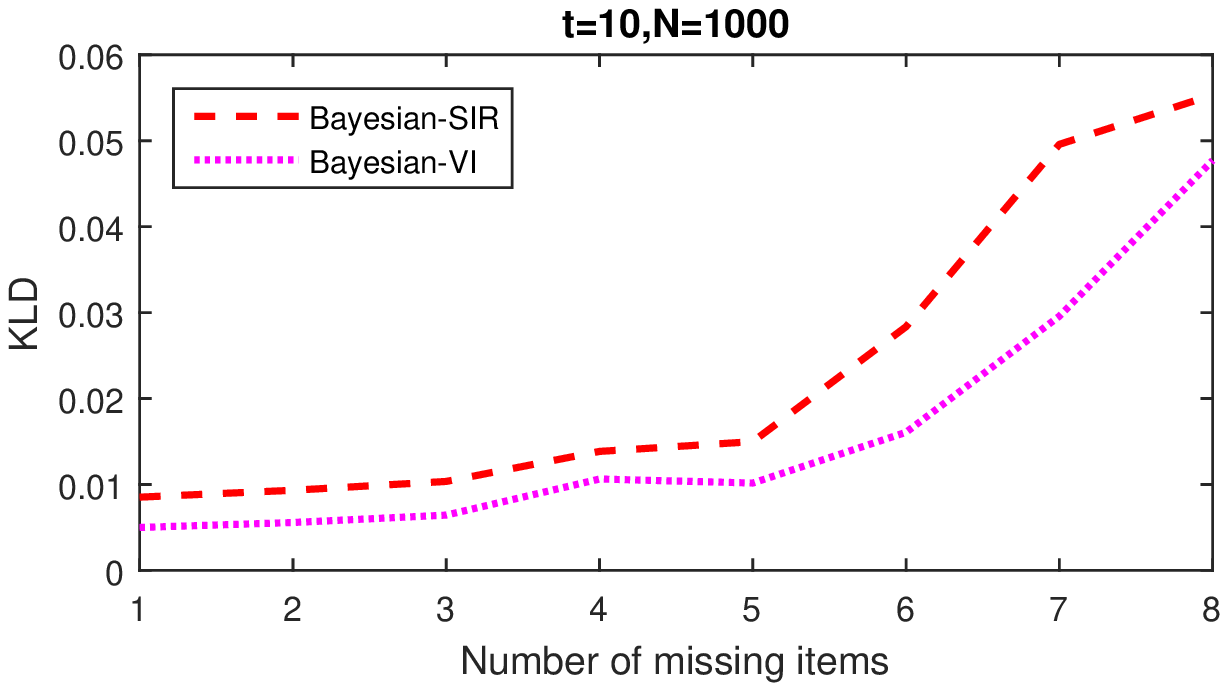}
\par\end{centering}
\begin{centering}
\includegraphics[height=4cm]{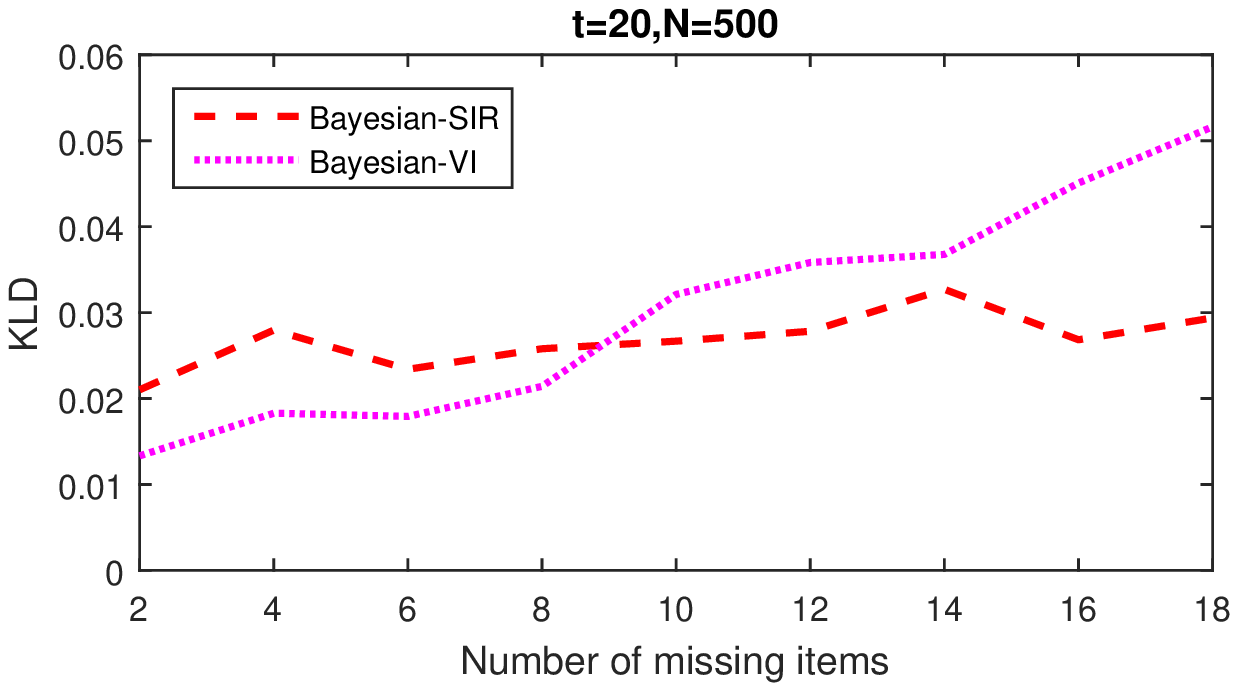}
\par\end{centering}
\caption{\label{fig:KLD-Incomplete_diff_N_miss}The KLD values between the
true model and the estimated model versus the number of missing items
with different settings: $t=10$, $N=500$ (top), $t=10$, $N=1000$
(middle) and $t=20$, $N=500$ (bottom). The estimated model is given
by Gibbs sampling with Bayesian method with SIR (Bayesian-SIR) and
the Bayesian method with variational approach (Bayesian-VI) in the
second conditional distribution. Each experiment is repeated 10 times
and the average results are shown. }
\end{figure}

\noindent The following experiments aim to compare the performance
with respect to the number of missing items $k$ when incomplete rankings
are observed. We first simulated data from our model with $\kappa=1$
and $\left\Vert \boldsymbol{\theta}\right\Vert =1$. Then we randomly
dropped the ranking for $k$ items and re-ranked the remaining items
to get the incomplete ranking. We chose three different settings for
the simulations: $\left(t=10,N=500\right)$, $\left(t=10,N=1000\right)$
and $\left(t=20,N=500\right)$. The number of missing items $k$ varies
up to $\left(t-2\right)$. We also calculated the Kullback-Leibler
divergence (KLD) of the true model from the estimated model to assess
the impact. The estimated model is given by Gibbs samplings with the
Bayesian method with SIR (Bayesian-SIR) and the Bayesian-VI in the
second conditional distribution. For each iteration for the Bayesian-SIR
method, we simulated $10$ samples from $100$ candidates and selected
one for the next step. The result of this comparison is shown in Figure
\ref{fig:KLD-Incomplete_diff_N_miss}. Each experiment is repeated
10 times and the average results are shown to smooth out the effect
of random initialization. 

From Figure \ref{fig:KLD-Incomplete_diff_N_miss}, we see that the
KLD increases with increasing number of missing items as expected.
From the comparison, the Bayesian-VI has lower KLD values for small
number of missing items $k$ compared to the Bayesian-SIR method.
This is consistent with previous simulation results. When $N$ is
large ($N=1,000$), Bayesian-VI performs better than Bayesian-SIR.
When the number of missing items is large, Bayesian-SIR seems to be
a better choice than Bayesian-VI. However, when comparing computation
time, Bayesian-VI is much faster than Bayesian-SIR.

\section{Applications}

\subsection{Sushi data sets}

\noindent We investigate the two data sets of \citet{kamishima2003nantonac}
for finding the difference in food preference patterns between eastern
and western Japan. Historically, western Japan has been mainly affected
by the culture of the Mikado emperor and nobles, while eastern Japan
has been the home of the Shogun and Samurai warriors. Therefore, the
preference patterns in food are different between these two regions
\citep{kamishima2003nantonac}. 

\begin{figure}[h]
\begin{centering}
\includegraphics[scale=0.7]{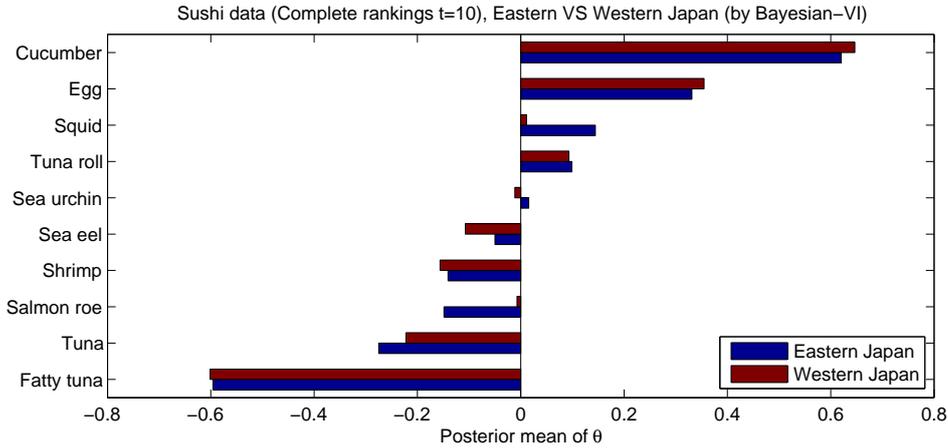}
\par\end{centering}
\caption{\label{fig:SUSHI_small_theta}Posterior means of $\boldsymbol{\theta}$
for the sushi complete ranking data ($t=10$) in Eastern Japan (Blue
bar) and Western Japan (Red bar) obtained by Bayesian-VI. }
\end{figure}

The first data set consists of complete rankings of $t=10$ different
kinds of sushi given by 5000 respondents according to their preference.
The region of respondents is also recorded ($N=3285$ for Eastern
Japan, $1715$ for Western Japan). We apply the MLE, Bayesian-SIR
and Bayesian-VI on both Eastern and Western Japan data. The settings
for the priors are similar to those used in the simulations in Section
\ref{subsec:Experiments-different_N}. Since the sample size $N$
is quite large compared to $t$, the estimated models for all three
methods are almost the same. Figure \ref{fig:SUSHI_small_theta} compares
the posterior means of $\boldsymbol{\theta}$ between Eastern Japan
(Blue bar) and Western Japan (Red bar) obtained by Bayesian-VI method.
 Note that the more negative value of $\theta_{i}$ means that the
more preferable sushi $i$ is. From Figure \ref{fig:SUSHI_small_theta},
we see that the main difference for sushi preference between Eastern
and Western Japan occurs in Salmon roe, Squid, Sea eel, Shrimp and
Tuna. People in Eastern Japan have a greater preference for Salmon
roe and Tuna than the western Japanese. On the other hand, the latter
have a greater preference for Squid, Shrimp and Sea eel. Table \ref{tab:Comparison-posterior-Sushi-small}
shows the posterior parameter obtained by Bayesian-VI. It can be seen
that the eastern Japanese are slightly more cohesive than western
Japanese since the posterior mean of $\kappa$ is larger.

\begin{table}
\begin{centering}
\begin{tabular}{c|c|c}
\hline 
{\footnotesize{}Posterior Parameter} & {\footnotesize{}Eastern Japan} & {\footnotesize{}Western Japan}\tabularnewline
\hline 
{\footnotesize{}$\beta$} & {\footnotesize{}1458.85} & {\footnotesize{}741.61}\tabularnewline
{\footnotesize{}$a$} & {\footnotesize{}18509.84} & {\footnotesize{}9462.70}\tabularnewline
{\footnotesize{}$b$} & {\footnotesize{}3801.57} & {\footnotesize{}2087.37}\tabularnewline
\hline 
{\footnotesize{}Posterior Mean of $\kappa$} & {\footnotesize{}4.87} & {\footnotesize{}4.53}\tabularnewline
\hline 
\end{tabular}
\par\end{centering}
\caption{\label{tab:Comparison-posterior-Sushi-small}Posterior parameters
for the sushi complete ranking data ($t=10$) in Eastern Japan and
Western Japan obtained by Bayesian-VI.}
\end{table}

\begin{figure}[h]
\begin{centering}
\includegraphics[scale=0.7]{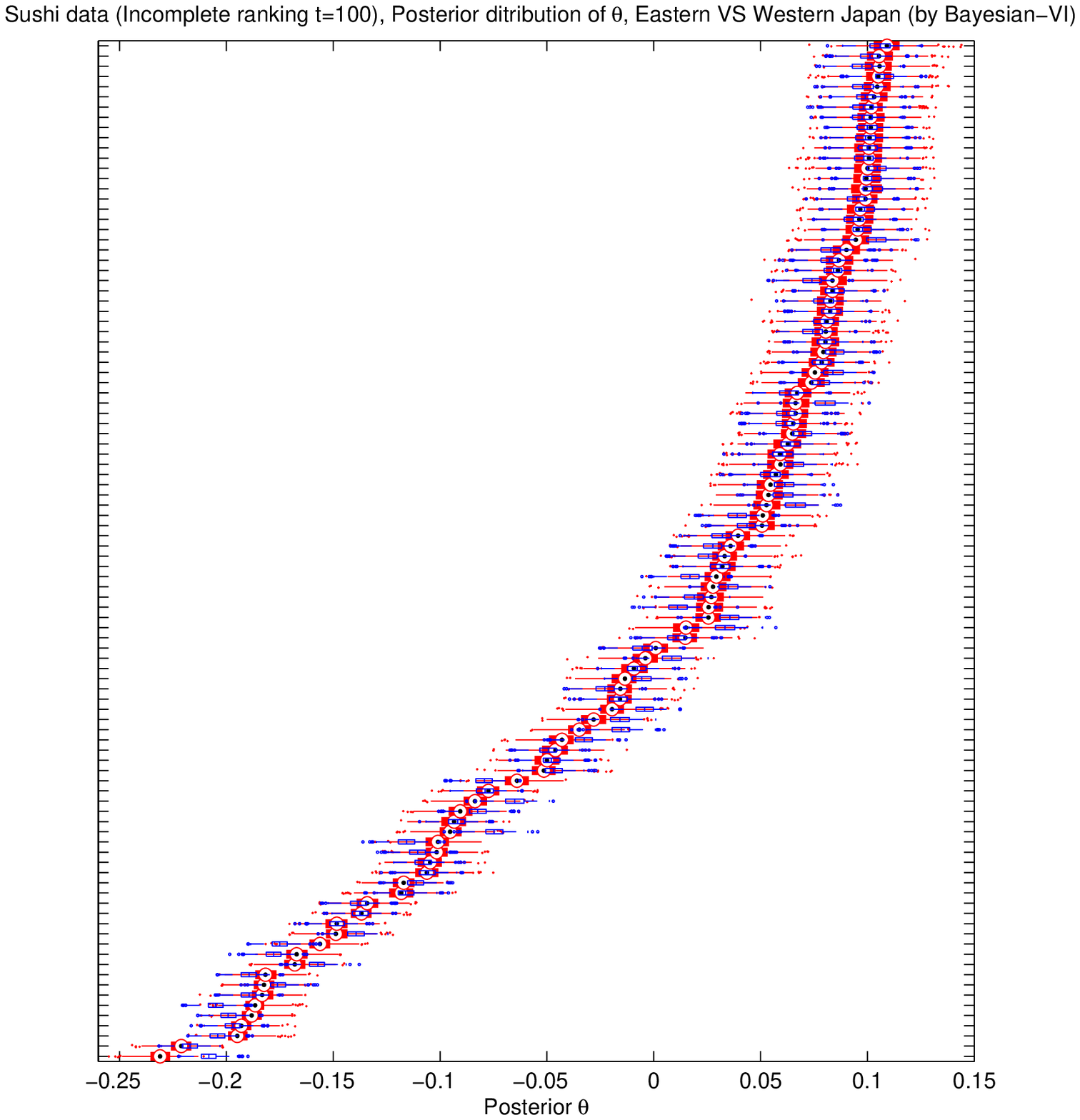}
\par\end{centering}
\caption{\label{fig:Comparison-boxplot-sushi-big}Boxplots of the posterior
means of $\boldsymbol{\theta}$ for the sushi incomplete rankings
($t=100$) in Eastern Japan (Blue box-plots) and Western Japan (Red
box-plots) obtained by Bayesian-VI.}
\end{figure}

The second data set contains incomplete rankings given by 5000 respondents
who were asked to pick and rank some of the $t=100$ different kinds
of sushi according to their preference and most of them only selected
and ranked the top 10 out of 100 sushi. Figure \ref{fig:Comparison-boxplot-sushi-big}
compares the box-plots of the posterior means of $\boldsymbol{\theta}$
between Eastern Japan (Blue box) and Western Japan (Red box) obtained
by Bayesian-VI. The posterior distribution of $\boldsymbol{\theta}$
is based on the Gibbs samplings after dropping the first 200 samples
during the burn-in period.  Since there are too many kinds of Sushi,
this graph doesn't allow us to show the name of each Sushi. However,
we can see that about one third of the 100 kinds of sushi have fairly
large posterior means of $\theta_{i}$ and their values are pretty
close to each others. This is mainly because these sushi are less
commonly preferred by Japanese and the respondents hardly chose these
sushi in their list. As these sushi are usually not ranked as top
10, it is natural to see that the posterior distributions of their
$\theta_{i}$'s tend to have a larger variance. 

\begin{figure}[h]
\begin{centering}
\includegraphics[scale=0.5]{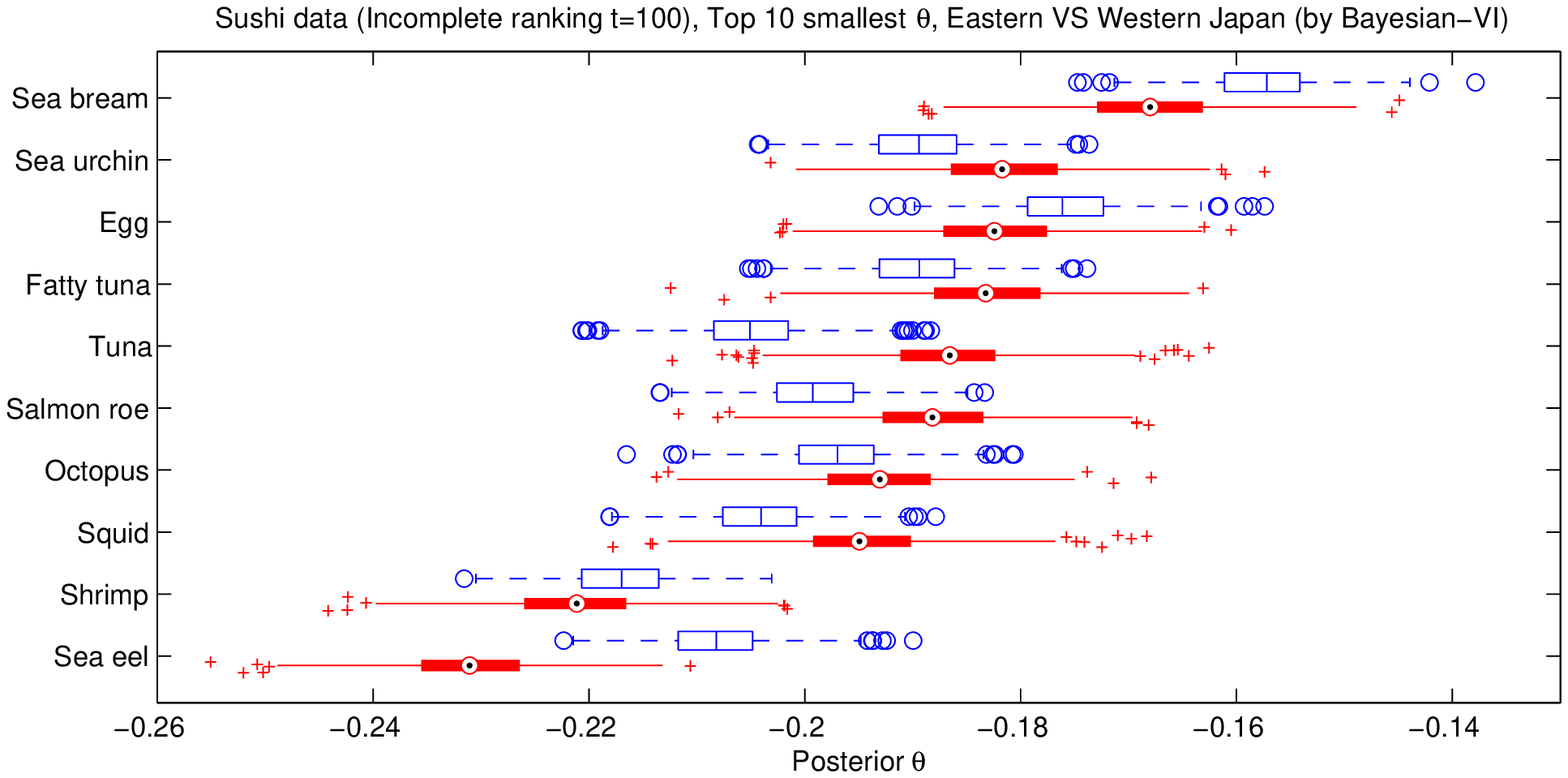}
\par\end{centering}
\caption{\label{fig:Comparison-boxplot-sushi-big-part}Box-plots of the top
10 smallest posterior means of $\boldsymbol{\theta}$ for the sushi
incomplete rankings ($t=100$) in Eastern Japan (Blue box-plots and
blue circles for outliers) and Western Japan (Red box-plots and red
pluses for outliers) obtained by Bayesian-VI.}
\end{figure}

From Figure \ref{fig:Comparison-boxplot-sushi-big}, we see that there
exists a greater difference between eastern and western Japan for
small $\theta_{i}$'s. Figure \ref{fig:Comparison-boxplot-sushi-big-part}
compares the box-plots of the top 10 smallest posterior means of $\boldsymbol{\theta}$
between Eastern Japan (Blue box) and Western Japan (Red box). The
main difference for sushi preference between Eastern and Western Japan
appears to be in Sea eel, Salmon roe, Tuna, Sea urchin and Sea bream.
The eastern Japanese prefer Salmon roe, Tuna and Sea urchin sushi
more than the western Japanese, while the latter like Sea eel and
Sea bream more than the former. Generally speaking, Tuna and Sea urchin
are more oily food, while Salmon roe and Tuna are more seasonal food.
So from the analysis of both data sets, we can conclude that the eastern
Japanese usually prefer more oily and seasonal food than the western
Japanese \citep{kamishima2003nantonac}.

\subsection{APA data}

\noindent We revisit the well-known APA data set of \citet{diaconis1988group}
which contains $5738$ full rankings of 5 candidates for the presidential
election of the American Psychological Association (APA) in 1980.
For this election, members of APA had to rank five candidates \{A,B,C,D,E\}
in order of their preference. Candidates A and C are research psychologists,
candidates D and E are clinical psychologists and candidate B is a
community psychologist. This data set has been studied by \citet{diaconis1988group}
and \citet{kidwell2008visualizing} who found that the voting population
was divided into 3 clusters. 

We fit the data using the mixture model stated in Section \ref{subsec:Mixture-ranking-model}.
We chose a non-informative prior for the Bayesian-VI method for a
different number of clusters $G=1$ to 5. Specifically, the prior
parameter $\boldsymbol{m}_{0g}$ is a randomly chosen unit vector
whereas $\beta_{0g}$, $d_{0g}$, $a_{0g}$ and $b_{0g}$ are chosen
as random numbers close to zero. The $p_{ig}$ are initialized as
$\frac{1}{G}$. Table \ref{tab:DIC-(Deviance-information} shows the
Deviance information criterion (DIC) for $G=1$ to $5$. It can be
seen that the mixture model with $G=3$ clusters attains the smallest
DIC. 

\begin{table}[h]
\begin{centering}
\begin{tabular}{c|ccccc}
\hline 
$G$ & 1 & 2 & 3 & 4 & 5\tabularnewline
\hline 
DIC & 54827 & 53497 & 53281 & 53367 & 53375\tabularnewline
\hline 
\end{tabular}
\par\end{centering}
\caption{\label{tab:DIC-(Deviance-information}Deviance information criterion
(DIC) for the APA ranking data.}
\end{table}

\begin{figure}[h]
\begin{centering}
\includegraphics[scale=0.8]{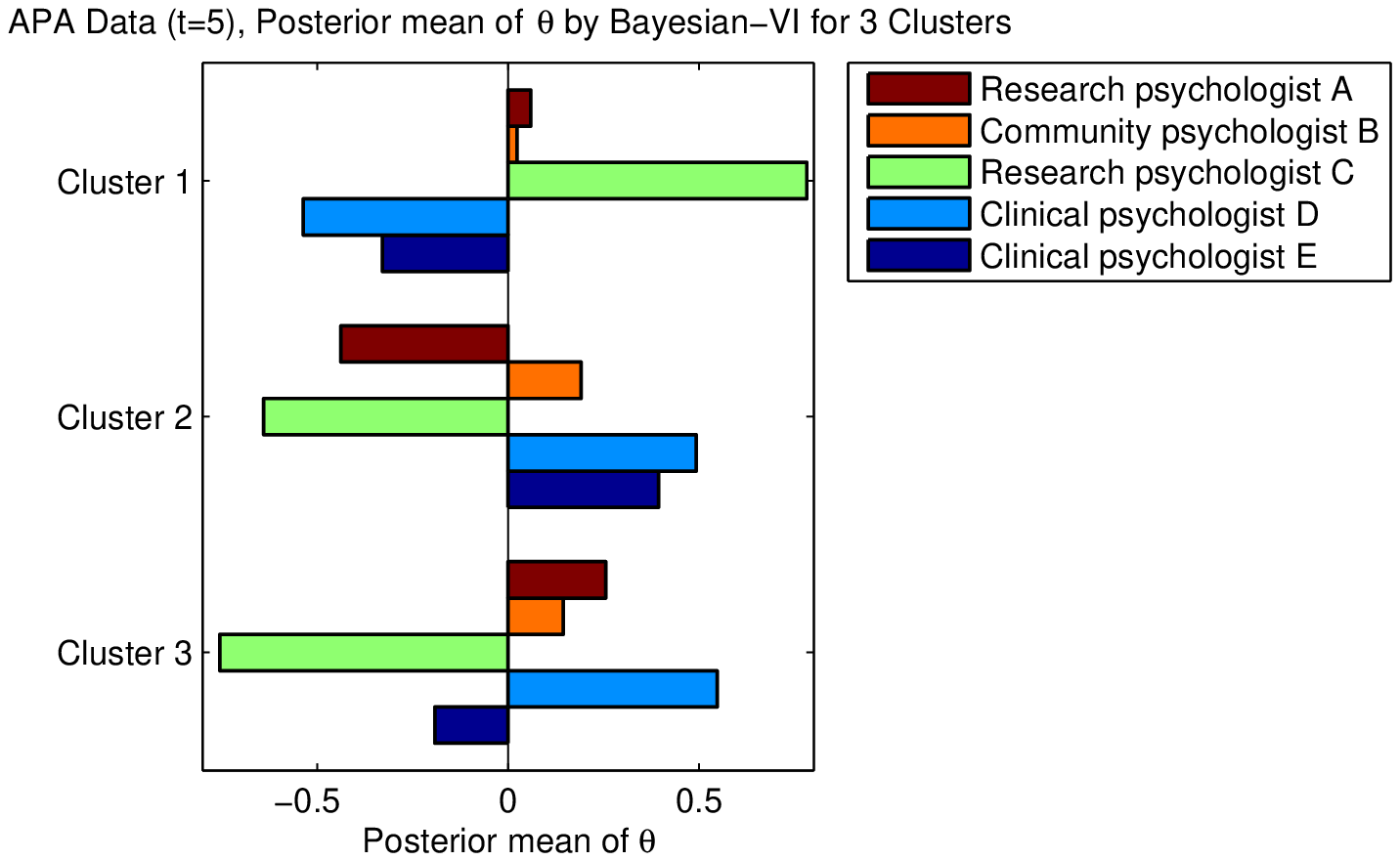}
\par\end{centering}
\caption{\label{fig:Comparison-APA-Data}Plot of the posterior means of $\boldsymbol{\theta}$
for the APA ranking data ($t=5$) for three clusters obtained by Bayesian-VI.}
\end{figure}

Table \ref{tab:The-posterior-parameter-APA} indicates the posterior
parameters for the three-cluster solution and Figure \ref{fig:Comparison-APA-Data}
exhibits the posterior means of $\boldsymbol{\theta}$ for the three
clusters obtained by Bayesian-VI. It is very interesting to see that
Clusters 1 vote clinical psychologists D and E as their first and
second choices and dislike especially the research psychologist C.
Cluster 2 prefer research psychologists A and C but dislike the others.
Cluster 3 prefer research psychologist C. From Table \ref{tab:The-posterior-parameter-APA},
Cluster 1 represents the majority (posterior mean of $\tau_{1}=56.31\%$).
Cluster 2 is small but more cohesive since the posterior mean of $\kappa_{2}$
is larger. Cluster 3 has a posterior mean of $\tau_{3}=20.73\%$ and
$\kappa_{3}$ is $1.52$. The preferences of the five candidates made
by the voters in the three clusters are heterogeneous and the mixture
model enables us to draw further inference from the data. 

\begin{table}
\begin{centering}
\begin{tabular}{c|ccc}
\hline 
{\footnotesize{}Posterior Parameter} & {\footnotesize{}Cluster 1} & {\footnotesize{}Cluster 2} & {\footnotesize{}Cluster 3}\tabularnewline
\hline 
{\footnotesize{}$\boldsymbol{m}$} & {\footnotesize{}0.06} & {\footnotesize{}-0.44} & {\footnotesize{}0.26}\tabularnewline
 & {\footnotesize{}0.02} & {\footnotesize{}0.19} & {\footnotesize{}0.14}\tabularnewline
 & {\footnotesize{}0.78} & {\footnotesize{}-0.64} & {\footnotesize{}-0.75}\tabularnewline
 & {\footnotesize{}-0.54} & {\footnotesize{}0.49} & {\footnotesize{}0.55}\tabularnewline
 & {\footnotesize{}-0.33} & {\footnotesize{}0.39} & {\footnotesize{}-0.19}\tabularnewline
{\footnotesize{}$\beta$} & {\footnotesize{}1067.10} & {\footnotesize{}1062.34} & {\footnotesize{}414.74}\tabularnewline
{\footnotesize{}$d$} & {\footnotesize{}3231.09} & {\footnotesize{}1317.21} & {\footnotesize{}1189.72}\tabularnewline
{\footnotesize{}$a$} & {\footnotesize{}4756.33} & {\footnotesize{}9224.97} & {\footnotesize{}1821.73}\tabularnewline
{\footnotesize{}$b$} & {\footnotesize{}3330.45} & {\footnotesize{}1239.41} & {\footnotesize{}1197.80}\tabularnewline
\hline 
{\footnotesize{}Posterior mean of $\boldsymbol{\kappa}$} & {\footnotesize{}1.43} & {\footnotesize{}7.44} & {\footnotesize{}1.52}\tabularnewline
{\footnotesize{}Posterior mean of $\boldsymbol{\tau}$} & {\footnotesize{}56.31\%} & {\footnotesize{}22.96\%} & {\footnotesize{}20.73\%}\tabularnewline
\hline 
\end{tabular}
\par\end{centering}
\caption{\label{tab:The-posterior-parameter-APA}Posterior parameters for the
APA ranking data ($t=5$) for three clusters obtained by Bayesian-VI.}

\end{table}

\subsection{Breast cancer gene expressions data}

\noindent We apply our mixture model on a ranked mRNA expression data
set to classify patients into the sub-type of breast cancer. Similar
topics have also been studied by \citet{naume2007presence}. All the
raw data can be obtained from the Stanford Micro array Database (SMD)
(http://genome-www5.stanford.edu/). We downloaded the mRNA expression
data of 121 breast cancer patients who have two disease sub-type based
on their ER/PgR-status: Estrogen Receptor negative (ER-, 41 patients)
or positive (ER+, 80 patients). Our aim is to classify the breast
cancer patients into two sub-groups based on their ranked gene expressions
data for 96 genes ($t=96$). These 96 genes are selected from the
KEGG Estrogen signaling pathway (Kyoto Encyclopedia of Genes and Genomes:
hsa04915) (http://www.genome.jp/kegg/). We use the rankings of 96
normalized log 2-transformed gene expression ratios for the 121 patients
as our training data. 

\begin{figure}
\begin{centering}
\includegraphics[scale=0.5]{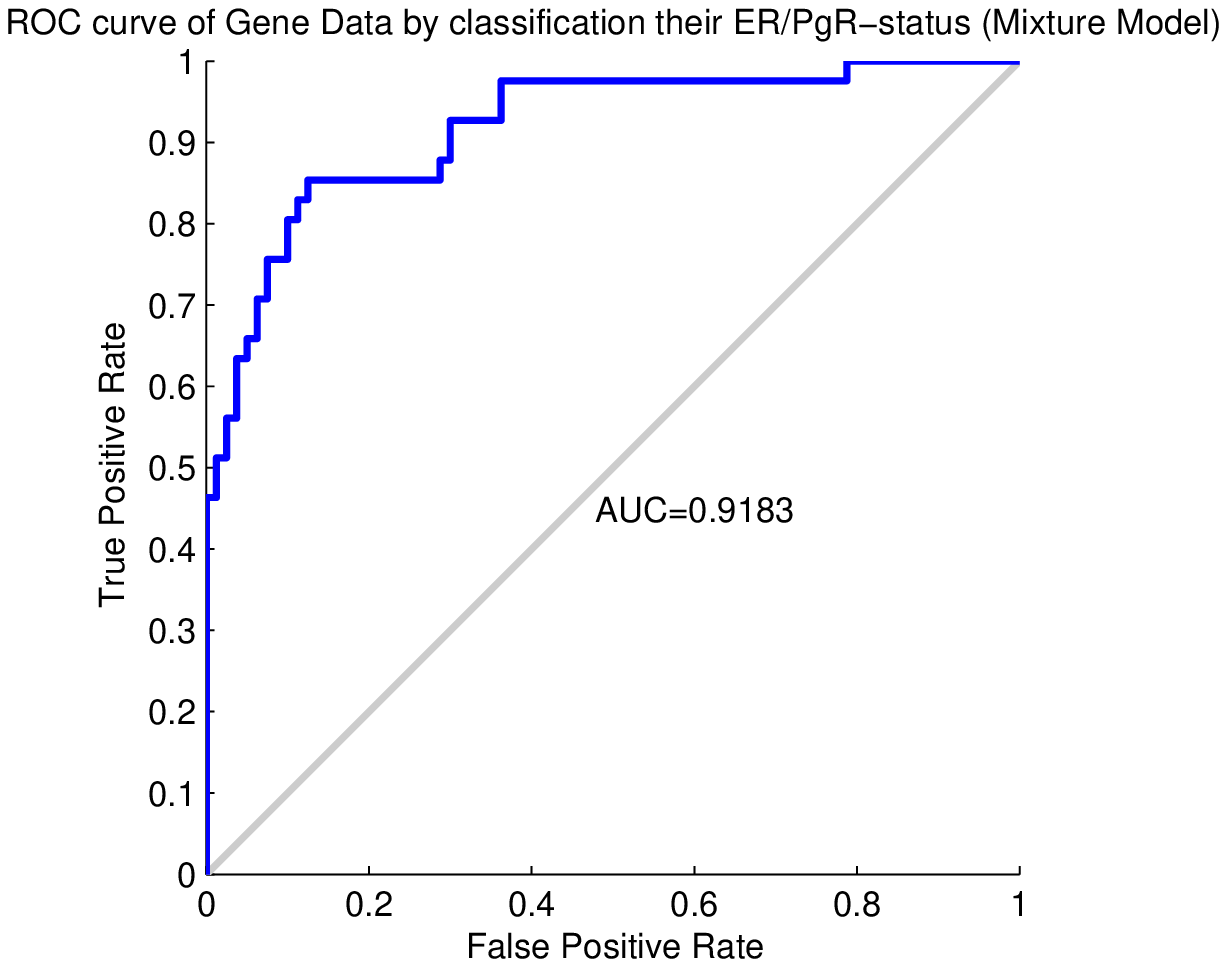}\includegraphics[scale=0.5]{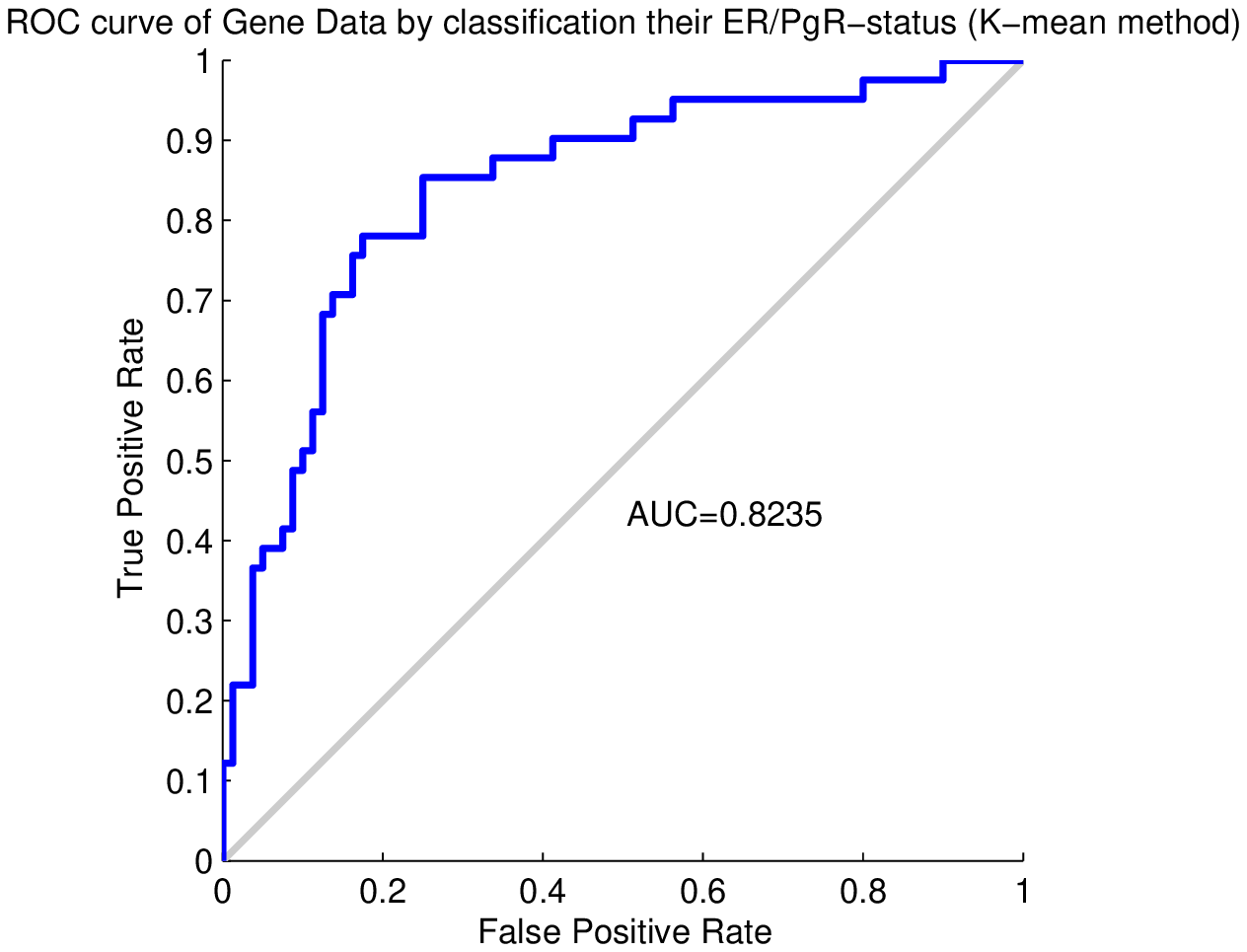}
\par\end{centering}
\caption{\label{fig:ROC-curve-for-Gene-data}ROC curves for classifying disease
sub type. The left panel is based on our fitted two-mixture model.
The right panel is based on the classification implied by K-means
clustering with squared Euclidean distance.}

\end{figure}

In this experiment, we first use the patients' gene ranking data (without
knowing the true disease sub-type of each patient) to fit our mixture
model ($G=2$). The prior parameter $\boldsymbol{m}_{0g}$ is a randomly
chosen unit vector while the other prior parameters $\beta_{0g}$,
$d_{0g}$, $a_{0g}$ and $b_{0g}$ are chosen as random small numbers
close to zero. The $p_{ig}$ are initialized as $\frac{1}{G}$. Table
\ref{tab:The-posterior-parameter-Gene-data} shows the posterior parameters
for the gene ranking data for the two clusters obtained by Bayesian-VI.
As the ER+ patients are more frequent in this data set, we label Cluster
1 as the ER+ group since the posterior mean of $\tau_{1}$ is higher
(66.79\%). So Cluster 2 is then labeled as the ER- group. Using our
clustering solution and the true disease sub-type for the patients,
Figure \ref{fig:ROC-curve-for-Gene-data} shows the ROC (Receiver
operating characteristic) curves based on the fitted two-mixture model
(the left panel) and the classification implied by the K-means clustering
with squared Euclidean distance (the right panel) \citep{hartigan1979algorithm,arthur2007k}.
From Figure \ref{fig:ROC-curve-for-Gene-data}, it is seen that our
mixture model has a greater discrimination power. The AUC (Area under
the curve) for our method is 0.9183 which is higher than that for
the K-means method (0.8235).

\begin{table}
\begin{centering}
\begin{tabular}{c|cc}
\hline 
{\footnotesize{}Posterior Parameter} & {\footnotesize{}Cluster 1 (ER+)} & {\footnotesize{}Cluster 2 (ER-)}\tabularnewline
\hline 
{\footnotesize{}$\beta$} & {\footnotesize{}63.68} & {\footnotesize{}29.75}\tabularnewline
{\footnotesize{}$d$} & {\footnotesize{}80.84} & {\footnotesize{}40.18}\tabularnewline
{\footnotesize{}$a$} & {\footnotesize{}16181.18} & {\footnotesize{}6462.97}\tabularnewline
{\footnotesize{}$b$} & {\footnotesize{}83.26} & {\footnotesize{}42.28}\tabularnewline
\hline 
{\footnotesize{}Posterior mean of $\kappa$} & {\footnotesize{}194.34} & {\footnotesize{}152.85}\tabularnewline
{\footnotesize{}Posterior mean of $\tau$} & {\footnotesize{}0.6679} & {\footnotesize{}0.3320}\tabularnewline
\hline 
\end{tabular}
\par\end{centering}
\caption{\label{tab:The-posterior-parameter-Gene-data}The posterior parameter
of $\theta$ for the gene ranking data ($t=96$) for two clusters
using Bayesian-VI.}
\end{table}

\section{Conclusions and Discussion}

\noindent We proposed a new class of general exponential ranking model
called angle-based ranking models. The model assumed a consensus score
vector $\boldsymbol{\theta}$ where the rankings reflect the rank-order
preference of the items. The probability of observing a ranking is
proportional to the cosine of the angle from the consensus score vector.
Then we proposed a very good approximation for the normalizing constant
using the von Mises-Fisher distribution which can facilitate the computation
of fitting the model. Usually it is an NP-hard problem to find the
estimates of parameters for other classes of ranking models when $t$
is large. However, our model avoided this problem and can easily calculate
the estimate of our model. We made use of Bayesian variational inference
to approximate the posterior density as well as the predictive density.
This approach exhibited a great computational advantage compared to
traditional MCMC methods. One can also consider to use regularization
methods such as LASSO, Ridge and Elastic Net to overcome the potential
over-fitting problem, especially for large $t$. In fact, regularization
methods can be implemented via a Bayesian approach with suitably chosen
priors. For instance, LASSO in a regression problem can be viewed
as maximum a posterior method in a Bayesian framework using a Laplace
prior centered at zero. It is of interest to study regularization
for angle-based models and such interesting problem would be studied
in the future. 

Unlike distance-based models, the consensus score  vector $\boldsymbol{\theta}$
proposed exhibits detailed information on item preferences while distance-based
model only provide equal-spaced modal ranking. We applied the method
to sushi data, and concluded that certain types of sushi are seldom
eaten by the Japanese. 

Model extensions to incomplete rankings and mixture models were also
developed. Incomplete rankings often arise when the number of items
ranked is large. The use of compatible rankings makes it possible
to handle incomplete rankings such as top-$k$ rankings, and subsets
ranking. The mixture models can be used as a model-based clustering
tool for ranking data. 

Our consensus score vector $\boldsymbol{\theta}$ defined on a unit
sphere can be easily reparameterized to incorporate additional arguments
or covariates in the model. the judge-specific covariates could be
age, gender and income, and the item-specific covariates could be
prices, weights and brands, and the judge-item-specific covariates
could be some personal experience on using each phone or brand. Adding
those covariates into the model will greatly improve the power of
prediction of our model. We can also develop Bayesian inference methods
to facilitate the computation. This interesting problem will be deferred
to later papers.

\section*{Acknowledgments}

\noindent The authors are grateful to the referees for making useful
suggestions which improved the presentation of several aspects of
the manuscript. The research of Philip L.H. Yu and Mayer Alvo was
supported by a grant from the Research Grants Council of the Hong
Kong Special Administrative Region, China (Project No.17303515). Mayer
Alvo was also supported by the Natural Sciences and Engineering Research
Council of Canada OGP0009068.

\section*{Appendix A. Derivation of the approximation for normalizing constant
of our model}

\noindent Since $t!$ permutations lie on a sphere in $(t-1)$-space,
our model is very close to another exponential family distribution,
the von Mises-Fisher distribution which is defined on a unit sphere.
Consider a von Mises-Fisher distribution defined on a $(t-1)$-space,
its normalizing constant can be written as the integration on a unit
$(t-2)$-sphere:
\begin{equation}
V_{t-1}(\kappa)^{-1}=\frac{(2\pi)^{\frac{t-1}{2}}I_{\frac{t-3}{2}}(\kappa)}{\kappa^{\frac{t-3}{2}}}=\int_{\left\Vert \boldsymbol{x}\right\Vert =1}\exp\left\{ \kappa\boldsymbol{\theta}^{T}\boldsymbol{x}\right\} d\boldsymbol{x}.\label{eq:NC_vMF}
\end{equation}
Using a naive Monte Carlo integration, we have
\[
\int_{\left\Vert \boldsymbol{x}\right\Vert =1}\exp\left\{ \kappa\boldsymbol{\theta}^{T}\boldsymbol{x}\right\} dx\simeq S\frac{1}{n}\sum_{i=1}^{n}\exp\left\{ \kappa\boldsymbol{\theta}^{T}\boldsymbol{x}_{i}\right\} ,
\]
where $\left\{ x_{i}\right\} $ are uniformly distributed on a $(t-2)$-unit
sphere, and $S=\int_{\left\Vert \boldsymbol{x}\right\Vert =1}d\boldsymbol{x}$. 

Summing over all possible $t!$ permutations $y_{i}$ we can further
write:
\begin{equation}
\int_{\left\Vert \boldsymbol{x}\right\Vert =1}\exp\left\{ \kappa\boldsymbol{\theta}^{T}\boldsymbol{x}\right\} d\boldsymbol{x}\simeq S\frac{1}{t!}\sum_{i=1}^{t!}\exp\left\{ \kappa\boldsymbol{\theta}^{T}\boldsymbol{y}_{i}\right\} .\label{eq:Approxi_NC_VMF}
\end{equation}
Note that
\[
S=\int_{\left\Vert \boldsymbol{x}\right\Vert =1}d\boldsymbol{x}=\frac{2\pi^{\frac{t-1}{2}}}{\Gamma(\frac{t-1}{2})}
\]
 is actually the surface of the unit $(t-2)$-sphere. After combining
(\ref{eq:NC_vMF}) and (\ref{eq:Approxi_NC_VMF}), we can have the
inverse of the approximation for normalizing constant of our model:
\begin{align*}
C_{t}(\kappa)^{-1}=\sum_{\boldsymbol{y}\in\text{\textgreek{R}}_{t}}\exp\left\{ \kappa\boldsymbol{\theta}^{T}\boldsymbol{y}\right\}  & \simeq\frac{t!}{S}V_{t-1}(\kappa)^{-1}\\
 & =\frac{t!}{S}\cdot\frac{(2\pi)^{\frac{t-1}{2}}I_{\frac{t-3}{2}}(\kappa)}{\kappa^{\frac{t-3}{2}}}\\
 & =\frac{2^{\frac{t-3}{2}}t!I_{\frac{t-3}{2}}(\kappa)\Gamma(\frac{t-1}{2})}{\kappa^{\frac{t-3}{2}}}.
\end{align*}

Note that when $\kappa=0$, the $V_{t-1}(\kappa)^{-1}$ becomes the
surface of a unit $(t-2)$-sphere: $V_{t-1}(\kappa)^{-1}=S$. Then
the approximation for normalizing constant of our model becomes:$C_{t}(\kappa)\simeq\frac{S}{t!}S^{-1}=\frac{1}{t!}$,
which is equal to the exact normalizing constant of our model for
$\kappa=0$.

\section*{Appendix B. Detailed Derivation of the predictive density of our
model}

\noindent To obtain the predictive density of our model, we first
integrate (\ref{eq:posterior_predictiver_density}) over $\boldsymbol{\theta}$:
\begin{align*}
\int p(\tilde{\boldsymbol{y}}|\kappa,\boldsymbol{\theta})vMF(\boldsymbol{\theta}|\boldsymbol{m},\beta\kappa)d\boldsymbol{\theta} & =C_{t}(\kappa)V_{t}(\beta\kappa)\int\exp\left[\kappa\boldsymbol{\theta}^{T}\tilde{\boldsymbol{y}}+\beta\kappa\boldsymbol{m}^{T}\boldsymbol{\theta}\right]d\boldsymbol{\theta}\\
 & =C_{t}(\kappa)V_{t}(\beta\kappa)V_{t}(\kappa\eta(\tilde{\boldsymbol{y}}))^{-1}\int V_{t}(\kappa\eta(\tilde{\boldsymbol{y}}))\exp\left[\kappa\eta(\tilde{\boldsymbol{y}})\frac{\tilde{\boldsymbol{y}}^{T}+\beta\boldsymbol{m}^{T}}{\eta(\tilde{\boldsymbol{y}})}\boldsymbol{\theta}\right]d\boldsymbol{\theta}
\end{align*}
where $\eta(\tilde{\boldsymbol{y}})=\left\Vert \tilde{\boldsymbol{y}}+\beta\boldsymbol{m}\right\Vert $.
This involves integrating a vMF with mean direction $\tilde{\boldsymbol{y}}+\beta\boldsymbol{m}$
and concentration parameter $\kappa\eta(\tilde{\boldsymbol{y}})$.
Hence, we can replace the known normalizing constant for vMF as:
\begin{align}
\int p(\tilde{\boldsymbol{y}}|\kappa,\boldsymbol{\theta})vMF(\boldsymbol{\theta}|\boldsymbol{m},\beta\boldsymbol{\kappa})d\boldsymbol{\theta} & =C_{t}(\kappa)V_{t}(\beta\kappa)V_{t}(\kappa\eta(\tilde{\boldsymbol{y}}))^{-1}\nonumber \\
 & =h(\tilde{\boldsymbol{y}})l(\kappa)\kappa^{\frac{t-3}{2}},\label{eq:Intergration_result_theta}
\end{align}
where 

\[
h(\tilde{\boldsymbol{y}})=\frac{1}{\Gamma\left(\frac{t-1}{2}\right)t!2^{\frac{t-3}{2}}}\left(\frac{\beta}{\eta(\tilde{\boldsymbol{y}})}\right)^{\frac{t-2}{2}},
\]
\[
l(\kappa)=\frac{I_{\frac{t-2}{2}}(\eta(\tilde{\boldsymbol{y}})\kappa)}{I_{\frac{t-3}{2}}(\kappa)I_{\frac{t-2}{2}}(\beta\kappa)}.
\]

Substituting (\ref{eq:Intergration_result_theta}) into $q(\tilde{\boldsymbol{y}}|\boldsymbol{Y})$
in (\ref{eq:posterior_predictiver_density}), we have
\begin{equation}
q\left(\tilde{\boldsymbol{y}}|\boldsymbol{Y}\right)=h(\tilde{\boldsymbol{y}})\frac{b^{a+\frac{t-1}{2}-1}}{\Gamma(a+\frac{t-1}{2}-1)}\int l(\kappa)e^{-b\kappa}\kappa^{a+\frac{t-1}{2}-2}d\kappa.\label{eq:posterior_preditive_density_one_intergration}
\end{equation}
Since the term $l(\kappa)$ involves three Bessel functions, we can
use a second order approximation of $\ln l(\kappa)$ in terms of $\kappa$
and $\ln\kappa$ as:
\begin{equation}
\ln l(\kappa)\approx\ln l(\bar{\kappa})-r(\tilde{\boldsymbol{y}})\left(\kappa-\bar{\kappa}\right)+s(\tilde{\boldsymbol{y}})(\ln\kappa-\ln\bar{\kappa}),\label{eq:approxi_lgl_alpha}
\end{equation}
where $r(\tilde{\boldsymbol{y}})$ and $s(\tilde{\boldsymbol{y}})$
are calculated from the first and second order derivatives expanded
at $\bar{\kappa}$. 

This yields:
\[
s(\tilde{\boldsymbol{y}})=-\eta^{2}(\tilde{\boldsymbol{y}})\bar{\kappa}^{2}\left(\frac{I'_{\frac{t-2}{2}}(\eta(\tilde{\boldsymbol{y}})\bar{\kappa})}{I_{\frac{t-2}{2}}(\eta(\tilde{\boldsymbol{y}})\bar{\kappa})}\right)'+\beta^{2}\bar{\kappa}^{2}\left(\frac{I'_{\frac{t-2}{2}}(\beta\bar{\kappa})}{I_{\frac{t-2}{2}}(\beta\bar{\kappa})}\right)'+\bar{\kappa}^{2}\left(\frac{I'_{\frac{t-3}{2}}(\bar{\kappa})}{I_{\frac{t-3}{2}}(\bar{\kappa})}\right)',
\]
\[
r(\tilde{\boldsymbol{y}})=\frac{s(\tilde{\boldsymbol{y}})}{\bar{\kappa}}-\eta(\tilde{\boldsymbol{y}})\frac{I'_{\frac{t-2}{2}}(\eta(\tilde{\boldsymbol{y}})\bar{\kappa})}{I_{\frac{t-2}{2}}(\eta(\tilde{\boldsymbol{y}})\bar{\kappa})}+\beta\frac{I'_{\frac{t-2}{2}}(\beta\bar{\kappa})}{I_{\frac{t-2}{2}}(\beta\bar{\kappa})}+\frac{I'_{\frac{t-3}{2}}(\bar{\kappa})}{I_{\frac{t-3}{2}}(\bar{\kappa})}.
\]
 The quantities $\frac{I'_{v}(x)}{I_{v}(x)}$ and $\left(\frac{I'_{v}(x)}{I_{v}(x)}\right)'$
can be computed using the recurrence relation of the derivative of
the modified Bessel function of the first kind:
\[
\frac{I'_{v}(x)}{I_{v}(x)}=\frac{I_{v+1}(x)}{I_{v}(x)}+\frac{v}{x}
\]
 
\[
\left(\frac{I'_{v}(x)}{I_{v}(x)}\right)'=-\frac{v}{x^{2}}+1-\frac{2v+1}{x}\left(\frac{I_{v+1}(x)}{I_{v}(x)}\right)-\left(\frac{I_{v+1}(x)}{I_{v}(x)}\right)^{2}.
\]
 Using (\ref{eq:approxi_lgl_alpha}), then the integration over $\kappa$
can be approximated by
\begin{align}
\int l(\kappa)e^{-b\kappa}\kappa^{a+\frac{t-1}{2}-2}d\kappa & \approx l(\bar{\kappa})e^{r(\tilde{\boldsymbol{y}})\bar{\kappa}}\bar{\kappa}^{-s(\tilde{\boldsymbol{y}})}\int e^{-\kappa\left(b+r(\tilde{\boldsymbol{y}})\right)}\kappa^{a+s(\tilde{\boldsymbol{y}})+\frac{t-1}{2}-2}d\kappa\nonumber \\
 & =l(\bar{\kappa})e^{r(\tilde{\boldsymbol{y}})\bar{\kappa}}\bar{\kappa}^{-s(\tilde{\boldsymbol{y}})}\Gamma\left(a+s(\tilde{\boldsymbol{y}})+\frac{t-1}{2}-1\right)\left(b+r(\tilde{\boldsymbol{y}})\right)^{-(a+s(\tilde{\boldsymbol{y}})+\frac{t-1}{2}-1)}\label{eq:result_intergration_over_alpha}
\end{align}
where the integration involves a Gamma distribution with shape parameter
\[
a+s(\tilde{\boldsymbol{y}})+\frac{t-1}{2}-1
\]
 and rate parameter 
\[
b+r(\tilde{\boldsymbol{y}}).
\]
 Hence, plugging in the known normalizing constant of the Gamma distribution,
we see that the approximate predictive density of $\tilde{\boldsymbol{y}}$
can be obtained by substituting (\ref{eq:result_intergration_over_alpha})
in (\ref{eq:posterior_preditive_density_one_intergration}):
\[
q(\tilde{\boldsymbol{y}}|\boldsymbol{Y})\approx h(\tilde{\boldsymbol{y}})l(\bar{\kappa})e^{r(\tilde{\boldsymbol{y}})\bar{\kappa}}\bar{\kappa}^{-s(\tilde{\boldsymbol{y}})}\frac{b^{a+\frac{t-1}{2}-1}\Gamma(a+s(\tilde{\boldsymbol{y}})+\frac{t-1}{2}-1)}{\left(b+r(\tilde{\boldsymbol{y}})\right)^{a+s(\tilde{\boldsymbol{y}})+\frac{t-1}{2}-1}\Gamma(a+\frac{t-1}{2}-1)}.
\]

\section*{Appendix C. Derivation of the variational inference of the mixture
ranking model}

\noindent For the mixture model, the evidence lower bound is given
by
\begin{equation}
\mathcal{L_{M}}(q)=E_{q(\boldsymbol{Z},\boldsymbol{\Theta},\boldsymbol{\kappa},\boldsymbol{\tau})}\left[\ln\frac{p(\boldsymbol{R}|\boldsymbol{Z},\boldsymbol{\Theta},\kappa)p(\boldsymbol{\Theta},\kappa)p(\boldsymbol{Z}|\tau)p(\tau)}{q(\boldsymbol{Z})q(\boldsymbol{\Theta}|\boldsymbol{\kappa})q(\boldsymbol{\kappa})q(\tau)}\right].\label{eq:evidence_lower_bound-mixture}
\end{equation}
Focusing first on terms involving $Z$, we have from (\ref{eq:evidence_lower_bound-mixture})
\begin{align*}
\mathcal{L}(q) & =E_{q(\boldsymbol{Z},\boldsymbol{\Theta},\boldsymbol{\kappa},\boldsymbol{\tau})}\left[\ln\left(p(\boldsymbol{R}|\boldsymbol{Z},\boldsymbol{\Theta},\boldsymbol{\kappa})p(Z|\boldsymbol{\tau})\right)\right]-E_{q(\boldsymbol{Z})}\left[\ln q(\boldsymbol{Z})\right]+constant\\
 & =\sum_{i=1}^{N}\sum_{g=1}^{G}E_{q(\boldsymbol{Z})}\left[z_{ig}\rho_{ig}\right]-E_{q(\boldsymbol{Z})}\left[\ln q(\boldsymbol{Z})\right]+constant,
\end{align*}
where 
\[
\rho_{ig}=\frac{t-3}{2}E_{q(\boldsymbol{\kappa})}(\ln\kappa_{g})+E_{q(\boldsymbol{\tau})}\left(\ln\tau_{g}\right)+E_{q(\boldsymbol{\Theta},\boldsymbol{\kappa})}\left(\kappa_{g}\boldsymbol{\theta}_{g}^{T}\boldsymbol{y}_{i}\right)-E_{q(\boldsymbol{\kappa})}\left(\ln I_{\frac{t-3}{2}}\left(\kappa_{g}\right)\right)-\ln\left[2^{\frac{t-3}{2}}t!\Gamma\left(\frac{t-1}{2}\right)\right].
\]
Since the term $E_{q(\boldsymbol{\kappa})}\left(\ln I_{\frac{t-3}{2}}\left(\kappa_{g}\right)\right)$
is not tractable, we use the method in Section \ref{subsec:Optimization-of-the-VI}
which leads to the lower bound
\[
\mathcal{L}_{M}(q)\geq\underline{\mathcal{L}_{M}(q)}=\sum_{i=1}^{N}\sum_{g=1}^{G}E_{q(\boldsymbol{Z})}\left[z_{ig}\underline{\rho_{ig}}\right]-E_{q(\boldsymbol{Z})}\left[\ln q(\boldsymbol{Z})\right]+constant.
\]
Using (\ref{eq:Ineq-I}), we have 
\begin{align}
\rho_{ig} & \geq\underline{\rho_{ig}}=\frac{t-3}{2}E_{q(\boldsymbol{\kappa})}(\ln\kappa_{g})+E_{q(\boldsymbol{\tau})}\left(\ln\tau_{g}\right)+E_{q(\boldsymbol{\Theta},\boldsymbol{\kappa})}\left(\kappa_{g}\boldsymbol{\theta}_{g}^{T}\boldsymbol{y}_{i}\right)-\ln\left[2^{\frac{t-3}{2}}t!\Gamma\left(\frac{t-1}{2}\right)\right]\label{eq:lower_bound_rho}\\
 & -\ln I_{\frac{t-3}{2}}\left(\bar{\kappa}_{g}\right)-\left(\frac{\partial}{\partial\kappa_{g}}\ln I_{\frac{t-3}{2}}\left(\bar{\kappa}_{g}\right)\right)\left[E_{q(\boldsymbol{\kappa})}\kappa_{g}-\bar{\kappa}_{g}\right].\nonumber 
\end{align}
Hence the optimal variational posterior distribution for $Z$ is
\[
\ln q^{*}(\boldsymbol{Z})=\sum_{i=1}^{N}\sum_{g=1}^{G}z_{ig}\rho_{ig}+constant
\]
which is recognized as a multinomial distribution:
\[
q^{*}(\boldsymbol{Z})=\prod_{i=1}^{N}\prod_{g=1}^{G}p_{ig}^{z_{ig}},
\]
where 
\[
p_{ig}=\frac{\exp(\underline{\rho_{ig}})}{\sum_{j=1}^{G}\exp(\underline{\rho_{ij}})}.
\]
Next, consider the optimization of $q(\tau)$. Since $E_{Z}(z_{ig})=p_{ig}$,
the optimal posterior distribution for $\tau$ can be written as 
\[
\ln q^{*}(\boldsymbol{\tau})=\sum_{g=1}^{G}\left(d_{0,g}-1+\sum_{i=1}^{N}p_{ig}\right)\ln\tau_{g}+constant,
\]
which is recognized to be a Dirichlet distribution with parameter
$d_{g}$:
\[
q^{*}(\boldsymbol{\tau})=Dirichlet(\boldsymbol{\tau}|\boldsymbol{d}),
\]
where $\boldsymbol{d}=\left[d_{1},...,d_{G}\right]^{T}$ and 
\begin{equation}
d_{g}=d_{0,g}+\sum_{i=1}^{N}p_{ig}.\label{eq:update_d_g-1}
\end{equation}

The remaining optimization of $q(\boldsymbol{\theta}|\kappa)$ and
$q(\kappa)$ is similar to Section \ref{subsec:Optimization-of-the-VI}
and we have 
\[
q^{*}(\boldsymbol{\theta}|\kappa)=\prod_{g=1}^{G}q^{*}(\boldsymbol{\theta}_{g}|\kappa_{g})
\]
 and 
\[
q^{*}(\boldsymbol{\theta}_{g}|\kappa_{g})=vMF(\boldsymbol{\theta}_{g}|\boldsymbol{m}_{g},\kappa_{g}\beta_{g}),
\]
where 
\begin{equation}
\beta_{g}=\left\Vert \beta_{0,g}\boldsymbol{m}_{0,g}+\sum_{i=1}^{n}p_{ig}\boldsymbol{y}_{i}\right\Vert ,\label{eq:Update_beta_g-1}
\end{equation}
\begin{equation}
\boldsymbol{m}_{g}=\left(\beta_{0,g}\boldsymbol{m}_{0,g}+\sum_{i=1}^{n}p_{ig}\boldsymbol{y}_{i}\right)\beta_{g}^{-1}.\label{eq:Update_m_g-1}
\end{equation}

We can write $q^{*}(\boldsymbol{\kappa})=\prod_{g=1}^{G}q^{*}(\kappa_{g})$
where

\[
q^{*}(\kappa_{g})=Gamma(\kappa_{g}|a_{g},b_{g}),
\]
and 
\begin{equation}
a_{g}=a_{0,g}+\left(\frac{t-3}{2}\right)\sum_{i=1}^{N}p_{ig}+\beta_{g}\bar{\kappa}_{g}\left[\frac{\partial}{\partial\beta_{g}\kappa_{g}}\ln I_{\frac{t-2}{2}}(\beta_{g}\bar{\kappa}_{g})\right],\label{eq:update_postrerior_a-Mixture-1}
\end{equation}
\begin{equation}
b_{g}=b_{0,g}+\left(\sum_{i=1}^{N}p_{ig}\right)\frac{\partial}{\partial\kappa_{g}}\ln I_{\frac{t-3}{2}}(\bar{\kappa}_{g})+\beta_{0,g}\left[\frac{\partial}{\partial\beta_{0,g}\kappa_{g}}\ln I_{\frac{t-2}{2}}(\beta_{0,g}\bar{\kappa}_{g})\right].\label{eq:Update_posterior_b-mixture-1}
\end{equation}

Since all the optimal variational posterior distributions are determined,
the expectations in (\ref{eq:lower_bound_rho}) can be easily evaluated
by the property of $q^{*}$: 
\[
E_{q(\boldsymbol{\kappa})}(\ln\kappa_{g})=\psi(a_{g})-\ln(b_{g}),E_{q(\boldsymbol{\tau})}\left(\ln\tau_{g}\right)=\psi(d_{g})-\psi\left(\sum_{g=1}^{G}d_{g}\right),
\]
 where $\psi(.)$ is the digamma function
\[
E_{q(\boldsymbol{\Theta},\boldsymbol{\kappa})}\left(\kappa_{g}\boldsymbol{\theta}_{g}^{T}\boldsymbol{y}_{i}\right)=\frac{a_{g}}{b_{g}}\boldsymbol{m}_{g}^{T}\boldsymbol{y}_{i}
\]
 and
\[
E_{q(\boldsymbol{\kappa})}\kappa_{g}=\frac{a_{g}}{b_{g}}.
\]

\section*{Appendix D. Additional simulations for Section 5.1}

\begin{figure}[!t]
\begin{centering}
\includegraphics[scale=0.5]{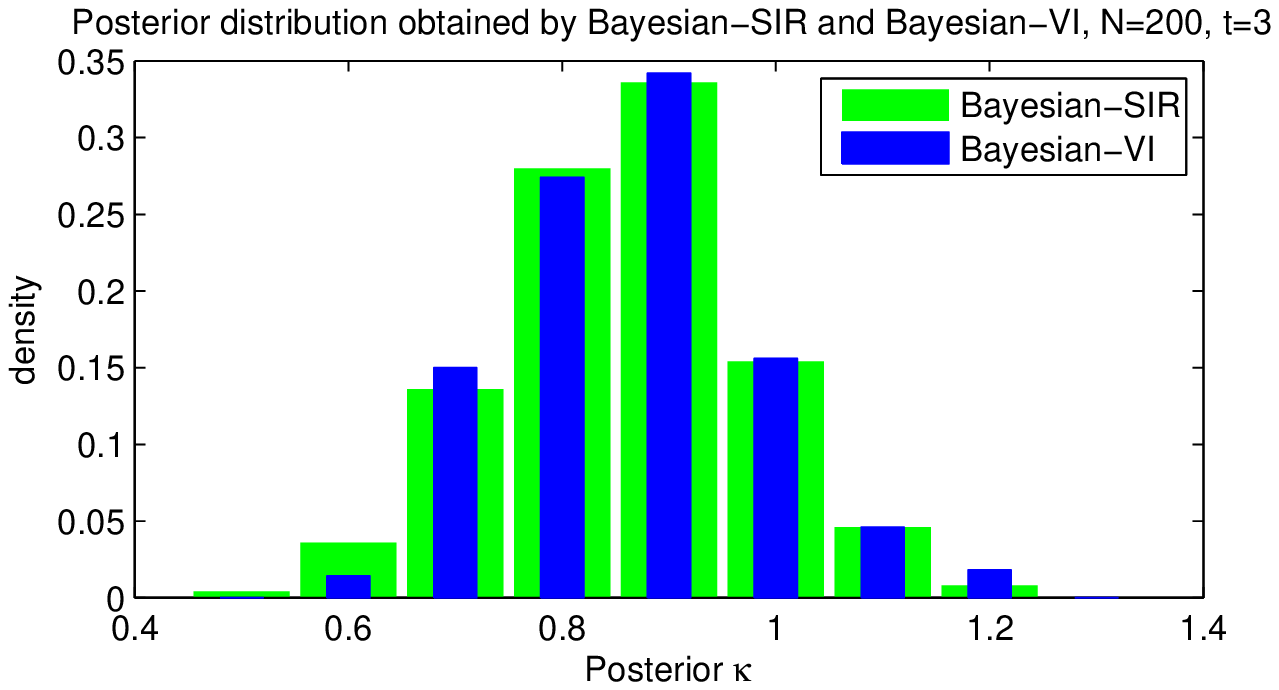}\includegraphics[scale=0.5]{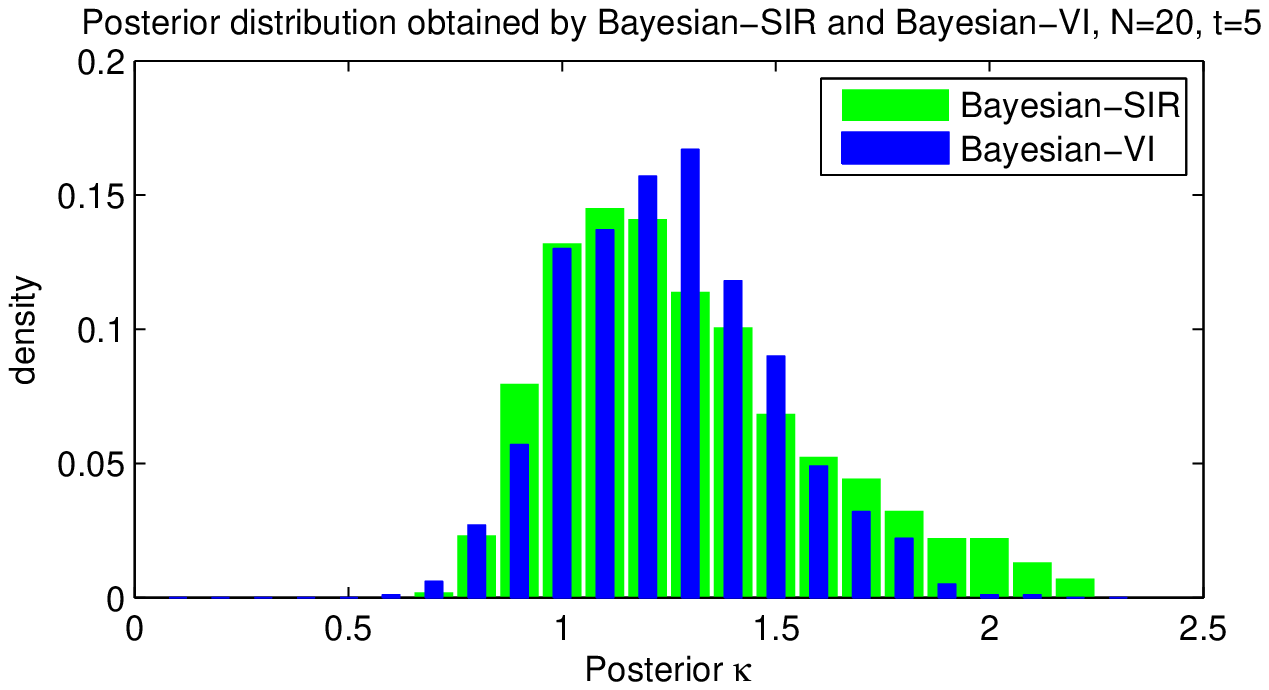}
\par\end{centering}
\begin{centering}
\includegraphics[scale=0.5]{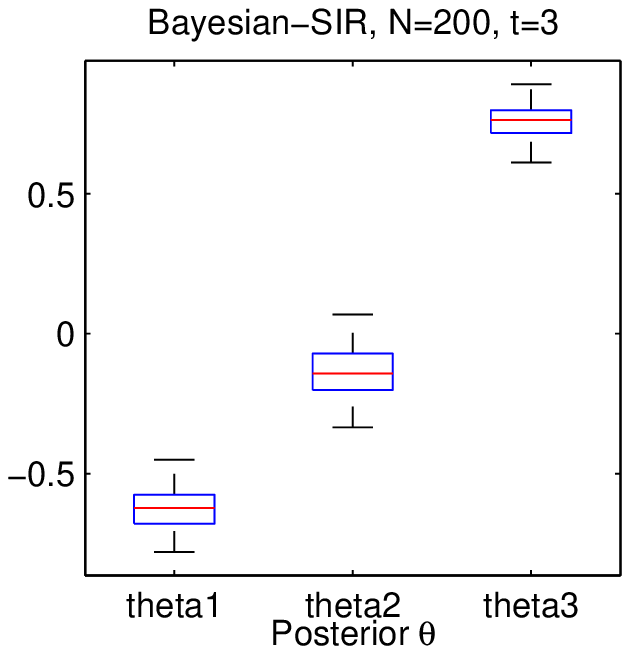}\includegraphics[scale=0.5]{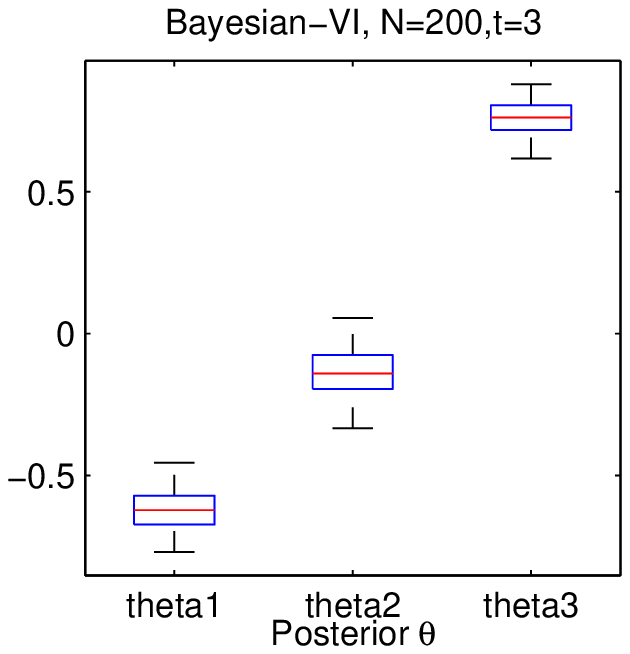}\includegraphics[scale=0.5]{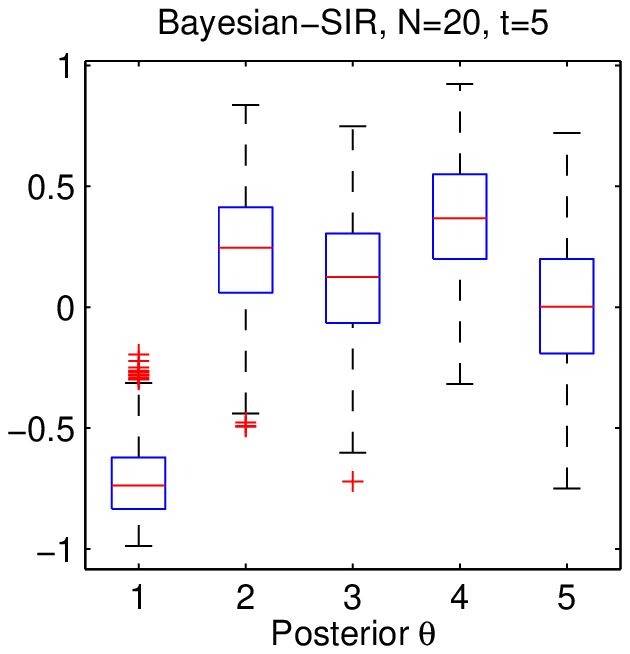}\includegraphics[scale=0.5]{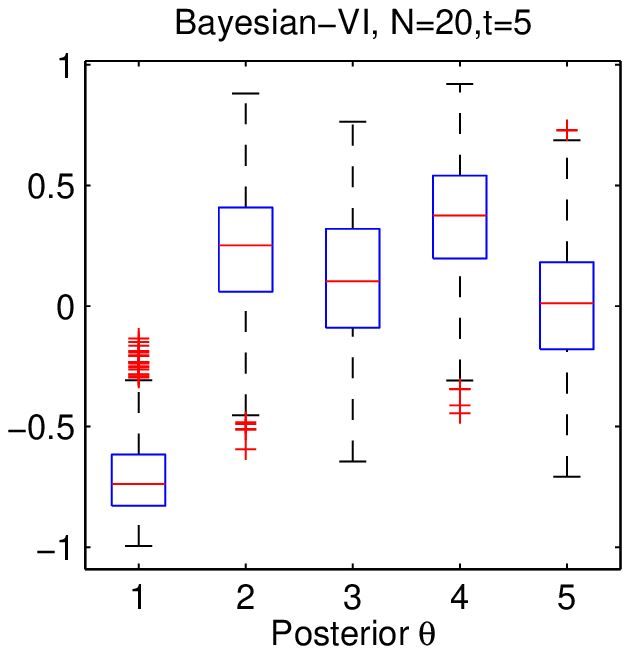}
\par\end{centering}
\begin{centering}
\includegraphics[scale=0.5]{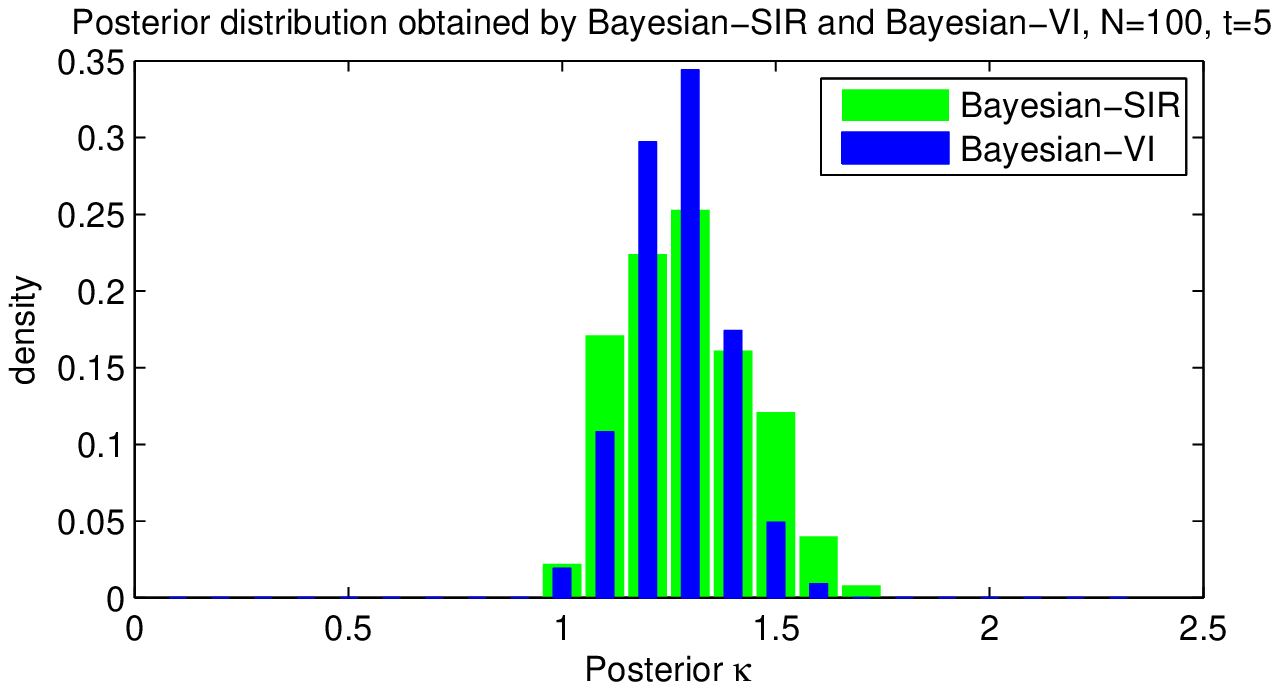}\includegraphics[scale=0.5]{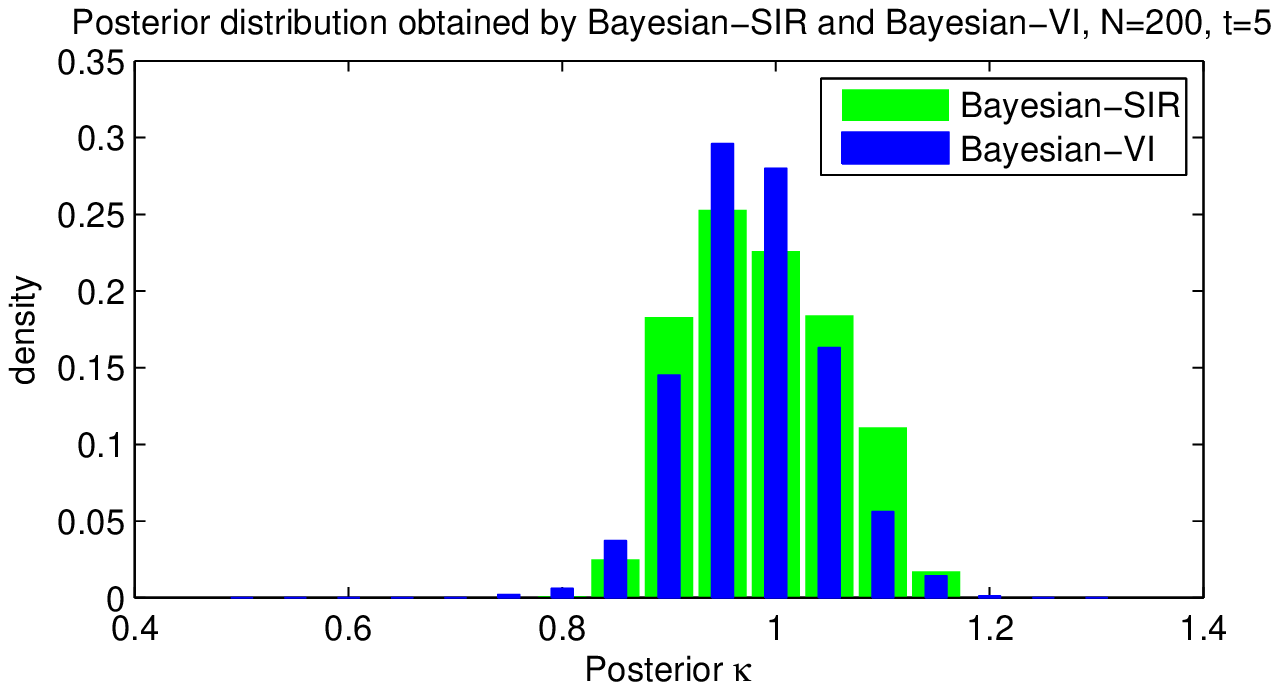}
\par\end{centering}
\begin{centering}
\includegraphics[scale=0.5]{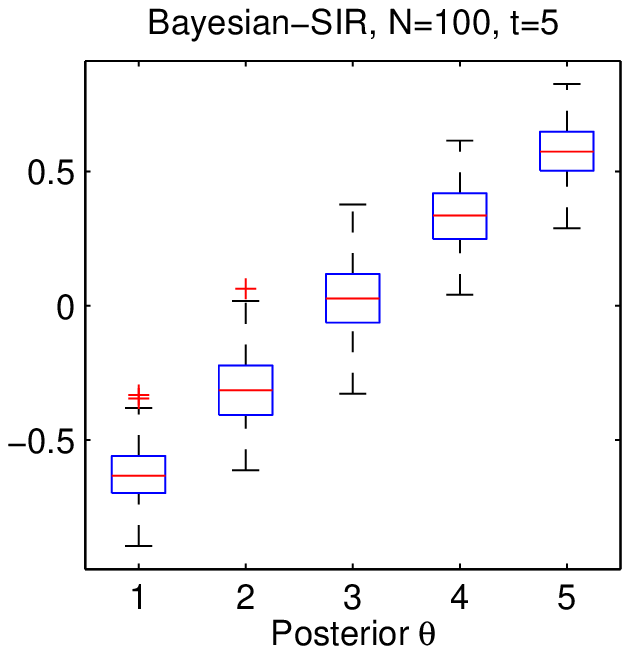}\includegraphics[scale=0.5]{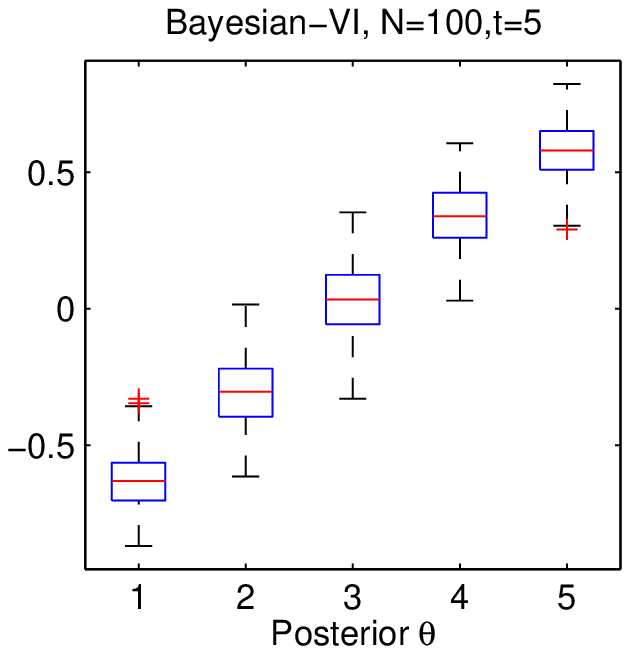}\includegraphics[scale=0.5]{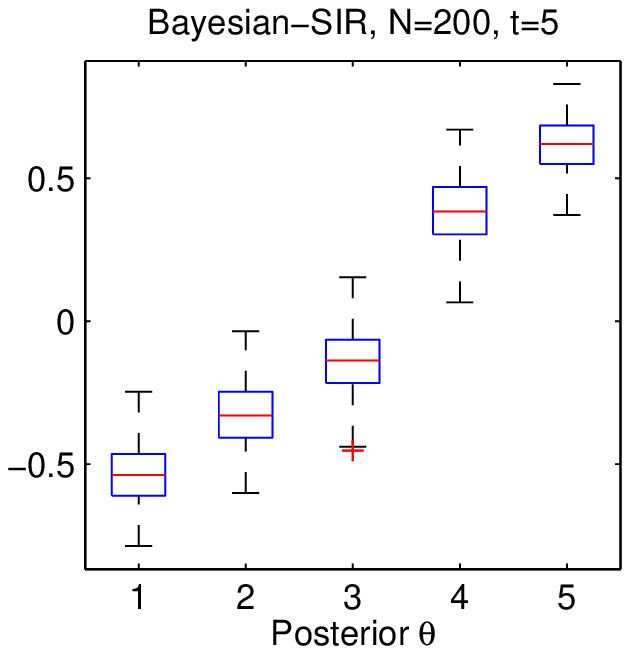}\includegraphics[scale=0.5]{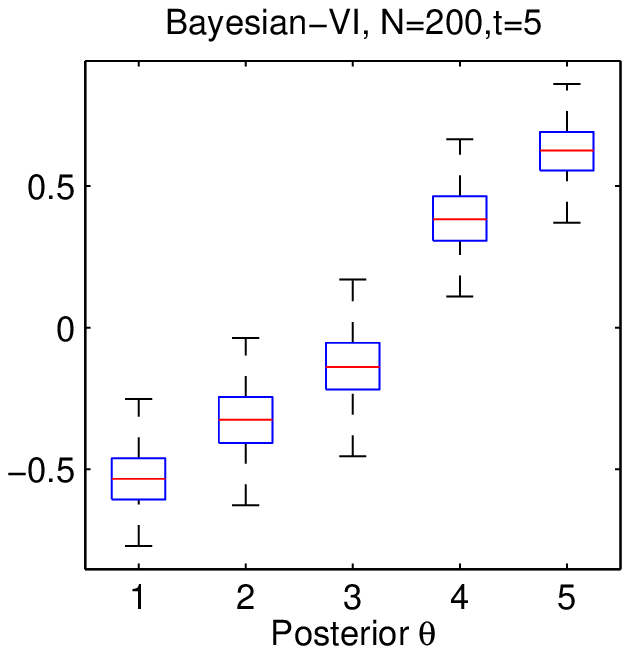}
\par\end{centering}
\caption{\label{fig:Comparison-posterior_distribution_SIR_VI-1}Comparison
of the posterior distribution obtained by Bayesian SIR method and
the approximate posterior distribution by variational inference approach.
The comparison is illustrated for different data sizes of $N=200,$
$t=3$ (top left) , $N=20$, $t=5$ (top right), $N=100,$ $t=5$
(bottom left) , $N=200$, $t=5$ (bottom right).}
\end{figure}

\noindent We have done more simulations to compare the true posterior
distribution with the approximate obtained using the variational inference
approach. We simulated another four data sets with $t=3,5$ and different
data sizes of $N=20,100,200.$ We generated samples from the posterior
distribution by SIR method in Section \ref{subsec:Bayesian-method-SIR}
using the proposal gamma density. We then applied the variational
approach in Algorithm \ref{alg:Bayesian-Estimation-our_model} and
generated samples from the corresponding posterior distribution. Figure
\ref{fig:Comparison-posterior_distribution_SIR_VI-1} exhibits the
histogram and box-plot for the posterior distribution of $\kappa$
and $\boldsymbol{\theta}$. From Figure \ref{fig:Comparison-posterior_distribution_SIR_VI-1},
we see that the posterior distribution using the Bayesian-VI is very
close to the posterior distribution obtained by the Bayesian-SIR method
for different cases of $t$ and $N$. 

\section*{References}

\bibliographystyle{apalike}
\bibliography{Bayesian_Ranking}

\begin{thebibliography}{}

\bibitem[Alvo and Cabilio, 1991]{alvo1991balanced}
Alvo, M. and Cabilio, P. (1991).
\newblock On the balanced incomplete block design for rankings.
\newblock {\em The Annals of Statistics}, 19(3):1597--1613.

\bibitem[Alvo and Yu, 2014]{alvo2014statistical}
Alvo, M. and Yu, P. L.~H. (2014).
\newblock {\em Statistical Methods for Ranking Data}.
\newblock Springer.

\bibitem[Arthur and Vassilvitskii, 2007]{arthur2007k}
Arthur, D. and Vassilvitskii, S. (2007).
\newblock k-means++: The advantages of careful seeding.
\newblock In {\em Proceedings of the eighteenth annual ACM-SIAM symposium on
  Discrete Algorithms}, pages 1027--1035. Society for Industrial and Applied
  Mathematics.

\bibitem[Banerjee et~al., 2005]{banerjee2005clustering}
Banerjee, A., Dhillon, I.~S., Ghosh, J., and Sra, S. (2005).
\newblock Clustering on the unit hypersphere using von {M}ises-{F}isher
  distributions.
\newblock {\em Journal of Machine Learning Research}, 6(Sep):1345--1382.

\bibitem[Blei et~al., 2017]{blei2017variational}
Blei, D.~M., Kucukelbir, A., and McAuliffe, J.~D. (2017).
\newblock Variational inference: A review for statisticians.
\newblock {\em Journal of the American Statistical Association},
  112(518):859--877.

\bibitem[Critchlow et~al., 1991]{critchlow1991probability}
Critchlow, D.~E., Fligner, M.~A., and Verducci, J.~S. (1991).
\newblock Probability models on rankings.
\newblock {\em Journal of Mathematical Psychology}, 35(3):294--318.

\bibitem[Diaconis, 1988]{diaconis1988group}
Diaconis, P. (1988).
\newblock {\em Group Representations in Probability and Statistics}.
\newblock Institute of Mathematical Statistics.

\bibitem[Fligner and Verducci, 1988]{Fligner1988}
Fligner, M. and Verducci, J.~S. (1988).
\newblock Multi-stage ranking models.
\newblock {\em Journal of the American Statistical Association}, 83:892--901.

\bibitem[Forbes and Mardia, 2015]{forbes2015fast}
Forbes, P.~G. and Mardia, K.~V. (2015).
\newblock A fast algorithm for sampling from the posterior of a von {M}ises
  distribution.
\newblock {\em Journal of Statistical Computation and Simulation},
  85(13):2693--2701.

\bibitem[Hartigan and Wong, 1979]{hartigan1979algorithm}
Hartigan, J.~A. and Wong, M.~A. (1979).
\newblock Algorithm {AS} 136: {A} k-means clustering algorithm.
\newblock {\em Journal of the Royal Statistical Society. Series C (Applied
  Statistics)}, 28(1):100--108.

\bibitem[Hornik and Gr{\"u}n, 2014]{hornik2014movmf}
Hornik, K. and Gr{\"u}n, B. (2014).
\newblock mov{MF}: An {R} package for fitting mixtures of von {M}ises-{F}isher
  distributions.
\newblock {\em Journal of Statistical Software}, 58(10):1--31.

\bibitem[Kamishima, 2003]{kamishima2003nantonac}
Kamishima, T. (2003).
\newblock Nantonac collaborative filtering: recommendation based on order
  responses.
\newblock In {\em Proceedings of the ninth ACM SIGKDD international conference
  on Knowledge Discovery and Data Mining}, pages 583--588. ACM.

\bibitem[Kidwell et~al., 2008]{kidwell2008visualizing}
Kidwell, P., Lebanon, G., and Cleveland, W. (2008).
\newblock Visualizing incomplete and partially ranked data.
\newblock {\em IEEE Transactions on Visualization and Computer Graphics},
  14(6):1356 -- 1363.

\bibitem[Lee and Yu, 2012]{lee2012mixtures}
Lee, P.~H. and Yu, P. L.~H. (2012).
\newblock Mixtures of weighted distance-based models for ranking data with
  applications in political studies.
\newblock {\em Computational Statistics \& Data Analysis}, 56(8):2486--2500.

\bibitem[Liu, 2008]{liu2008monte}
Liu, J.~S. (2008).
\newblock {\em Monte Carlo Strategies in Scientific Computing}.
\newblock Springer Science \& Business Media.

\bibitem[Mallows, 1957]{Mallows1957}
Mallows, C.~L. (1957).
\newblock Non-null ranking models. {I}.
\newblock {\em Biometrika}, 44(1/2):114--130.

\bibitem[Naume et~al., 2007]{naume2007presence}
Naume, B., Zhao, X., Synnestvedt, M., Borgen, E., Russnes, H.~G.,
  Lingj{\ae}rde, O.~C., Str{\o}mberg, M., Wiedswang, G., Kvalheim, G.,
  K{\aa}resen, R., et~al. (2007).
\newblock Presence of bone marrow micrometastasis is associated with different
  recurrence risk within molecular subtypes of breast cancer.
\newblock {\em Molecular Oncology}, 1(2):160--171.

\bibitem[Nunez-Antonio and Guti{\'e}rrez-Pena, 2005]{nunez2005bayesian}
Nunez-Antonio, G. and Guti{\'e}rrez-Pena, E. (2005).
\newblock A bayesian analysis of directional data using the von
  {M}ises--{F}isher distribution.
\newblock {\em Communications in Statistics-Simulation and Computation},
  34(4):989--999.

\bibitem[Sra, 2012]{sra2012short}
Sra, S. (2012).
\newblock A short note on parameter approximation for von {M}ises-{F}isher
  distributions: and a fast implementation of ${I}_s(x)$.
\newblock {\em Computational Statistics}, 27(1):177--190.

\bibitem[Taghia et~al., 2014]{taghia2014bayesian}
Taghia, J., Ma, Z., and Leijon, A. (2014).
\newblock Bayesian estimation of the von-{M}ises {F}isher mixture model with
  variational inference.
\newblock {\em IEEE Transactions on Pattern Analysis and Machine Intelligence},
  36(9):1701--1715.

\bibitem[Thurstone, 1927]{Thurstone1927a}
Thurstone, L.~L. (1927).
\newblock A law of comparative judgement.
\newblock {\em Psychological Reviews}, 34(4):273--286.

\bibitem[Yu, 2000]{yu2000bayesian}
Yu, P. L.~H. (2000).
\newblock Bayesian analysis of order-statistics models for ranking data.
\newblock {\em Psychometrika}, 65(3):281--299.

\bibitem[Yu et~al., 2005]{YuLamLo2005}
Yu, P. L.~H., Lam, K.~F., and Lo, S.~M. (2005).
\newblock Factor analysis for ranked data with application to a job selection
  attitude survey.
\newblock {\em Journal of the Royal Statistical Society Series A},
  168(3):583--597.

\end{thebibliography}

\end{document}